\documentclass[prd,twocolumn,showpacs,groupedaddress,superscriptaddress,amsmath,amssymb]{revtex4-2} 

\usepackage{hhline}
\usepackage{slashed}
\usepackage{graphicx}
\usepackage{graphics}
\usepackage{dcolumn}
\usepackage{bm}
\usepackage{amssymb}
\usepackage{enumerate}
\usepackage{lineno}
\usepackage{multirow}
\usepackage{float}

\def\u{\tilde{u}}
\def\s{\tilde{s}}
\def\t{\tilde{t}}
\def\P{\mathcal{P}}
\def\u{\tilde{u}}
\def\s{\tilde{s}}
\def\t{\tilde{t}}
\def\k{{\bf k}}
\def\p{{\bf p }}
\def\q{{\bf q }}

\begin{document}

\title{Probing Hidden Sectors with a muon beam:\\ Total and differential cross-sections for vector boson production in muon bremsstrahlung}

\author{D.~V.~Kirpichnikov}
\email[\textbf{e-mail}: ]{kirpich@ms2.inr.ac.ru}
\affiliation{Institute for Nuclear Research, 117312 Moscow, Russia}
\author{H. Sieber}
\thanks{Corresponding author}
\email[\textbf{e-mail}: ]{henri.hugo.sieber@cern.ch}
\affiliation{
ETH Z\"urich, Institute for Particle Physics and Astrophysics,
CH-8093 Z\"urich, Switzerland}
\author{L. Molina Bueno}
\email[\textbf{e-mail}: ]{laura.molina.bueno@cern.ch}
\affiliation{
ETH Z\"urich, Institute for Particle Physics and Astrophysics,
CH-8093 Z\"urich, Switzerland}
\affiliation{Instituto de Fisica Corpuscular (CSIC/UV),
Carrer del Catedrátic José Beltrán Martinez, 2, 46980 Paterna, Valencia, Spain}
\author{P. Crivelli}
\email[\textbf{e-mail}: ]{paolo.crivelli@cern.ch}
\affiliation{
ETH Z\"urich, Institute for Particle Physics and Astrophysics,
CH-8093 Z\"urich, Switzerland}
\author{M. M. Kirsanov}
\email[\textbf{e-mail}: ]{mikhail.kirsanov@cern.ch}
\affiliation{Institute for Nuclear Research, 117312 Moscow, Russia}

\date{\today}

\begin{abstract}
Vector bosons, such as Dark Photon $A'$ or  $Z'$, can couple to muons and be produced in 
the bremsstrahlung reaction $\mu^- + N \rightarrow \mu^- + N + A'(Z')$. Their possible 
subsequent invisible decay can be detected in fixed target experiments through missing 
energy/momentum signature. In such experiments, not only is the energy transfer to $A'(Z')$
important, but also the recoil muon angle $\psi_{\mu'}$. In this paper, we derive the total and 
the double differential cross-sections involved in this process using the phase space 
Weiszäcker-Williams and improved Weiszäcker-Williams approximations, as well as using exact-tree-level calculations. As an example, we compare the derived cross-sections and 
resulting signal yields in the NA64$\mu$ experiment that uses a 160 GeV muon beam at the 
CERN Super Proton Synchrotron accelerator. We also discuss its impact on the NA64$\mu$ expected sensitivity  to explore the $(g-2)_\mu$ anomaly favoured region with a $Z'$ boson considering $10^{12}$ muons accumulated on target. 
\end{abstract}

\maketitle

\section{Introduction}
The recently confirmed $4.2\sigma$ discrepancy in the anomalous magnetic moment measurement of the muon~\cite{Abi:2021gix} with respect to its theoretical prediction~\cite{Aoyama:2020ynm} 
\begin{equation}
     \Delta a_{\mu} \equiv a_{\mu}(exp) - a_{\mu}(th) =(251 \pm 59)\cdot 10^{-11}
\end{equation}
remains one of the long standing puzzles in the Standard Model (SM) of particle physics. 
A minimal extension of the SM able to explain this mismatch consists 
in the addition of a new weak interaction between the standard 
matter and a dark sector. The existence of dark sectors is also 
strongly motivated as a framework to explain the origin of Dark Matter
(DM) as a thermal freeze-out relic, a mechanism similar to the one of the Weakly 
Interacting Massive Particles (WIMPs), but in a broader and lower mass 
range \cite{Lanfranchi:2020crw}. In particular, the $L_\mu -L_\tau$ 
models based on the existence of a new light  vector boson ($Z'$) 
which predominantly couples to the second and third generation of 
leptons, are theoretically well motivated as they are an anomaly-free 
extension of the gauge sector. The interaction between $Z'$ and the 
$L_{\mu}-L_{\tau}$ current is given by the following
term~\cite{Gninenko:2014pea}:

\begin{equation}
    L_{Z'} = g'\left(\bar{\mu}\gamma_{\nu}\mu + \bar{\nu}_{\mu L}\gamma_{\nu}\nu_{\mu L}-\bar{\tau}\gamma_{\nu}\tau- \bar{\nu}_{\tau L}\gamma_{\nu}\nu_{\tau L} \right)Z'^{\nu},
\end{equation}

This interaction can give via loop effects the required additional contribution to explain the discrepancy between the theoretical and experimental value of the muon magnetic moment \cite{Gninenko:2001hx, Gninenko:2014pea, Chen:2017awl, Gninenko:2018tlp, Kirpichnikov:2020tcf, Amaral:2021rzw}:

\begin{equation}
    \Delta a^{Z'}_{\mu} = \frac{g'^2}{4\pi^2}\int_{0}^{1} dx \frac{x^2\left(1-x\right)}{x^2+\left(1-x\right)m_{Z'}^2/m^2_{\mu}},
\end{equation}
for couplings $g'\sim 10^{-3}-10^{-5}$ and $Z'$ masses in the MeV-GeV range.


If such a boson exists, it can be produced in the muon nuclear 
bremsstrahlung process $\mu^-+N \to \mu^-+N+Z'$ when a high 
energy muon beam impinges on a target. In this paper, we focus on the calculation of this production mechanism but the 
same results can be directly applied to the $A'$ bremsstrahlung ($A'$-strahlung) production from an initial muon beam. 
In order to illustrate the impact of our results, we consider the NA64$\mu$ experiment \cite{na64mu} which aims at 
exploring dark sector particles weakly coupled to muons. A similar experiment, $M^3$, is also under consideration at 
Fermilab \cite{Kahn:2018cqs}. Typically, the $Z’$ is expected to be produced in the bremsstrahlung reaction of a 160 GeV
muon beam impinging on an electromagnetic calorimeter acting as an active target. A fraction of the primary beam energy is 
carried away by the scattered muon, which momentum is measured by a set of trackers after being deflected by a magnet 
located downstream the target. 
 In this work, we derive the cross-sections involved in the production process and estimate its impact on the signal 
 yield using the Geant4-based \cite{Agostinelli:2002hh} dark matter package DMG4 \cite{Celentano:2021cna}, in which we 
 implemented the simulation of $Z’$ according to the derived cross-sections. We compare the results of analytical 
 calculations using phase space approximations such as Weiszäcker-Williams (WW) and improved Weiszäcker-Williams (IWW) with exact-tree-level (ETL) calculations.
 We also derive a new analytical expression to calculate the photon flux in the WW approximation. Finally, we 
 study the differential cross sections as a function of the recoil muon and $Z'$ angles.
 
 This paper is organised as follows. 
 In section \ref{ETLSection}, we calculate at ETL the total cross-section for $Z'$ production.
 In section \ref{WWCSSection}, we discuss the differential cross-sections for the angle
 and energy fraction of the emitted particle $Z'$ for both WW and IWW approximations. In 
 Section~\ref{WWCSSection}, we also discuss the zero mass limit $m_{Z'}\to 0$ of the 
 corresponding cross-sections and show 
 that WW and ETL reproduce the bremsstrahlung spectrum of muons on nucleus.
 In Section~\ref{MuonWWCSSection}, we derive the differential cross-sections for the angle
 and energy fraction of the deflected muon for both WW and IWW approximations.
  Finally, in section 
 \ref{BoundsSection}, we evaluate the impact of the cross-section calculations 
 on the projected $(g-2)_\mu$ sensitivities for NA64$_{\mu}$ experiment.
\section{The exact tree level calculation
\label{ETLSection}}
In this section, we discuss the derivation of the $Z'$-boson exact-tree-level production 
cross-section, in particular, we follow the notations of Ref.~\cite{Liu:2017htz, Gninenko:2017yus}. 
Let us consider the kinematic variables of the  process 
$$
\mu^-(p)+N(P_i)\rightarrow \mu^-(p')+N(P_f)+Z'(k),
$$
where $p=(E_\mu, {\bf p})$ is the four-momentum of the incoming muon, 
$P_i= (M, 0)$ designates the nucleus four-momentum in the laboratory frame, 
and $P_f= (P^0_f, {\bf P}_f )$ is the final state of the nucleus. The $Z'$-boson momentum is $k=(E_{Z'}, {\bf k})$ and $p'=(E_{\mu'}, {\bf p}')$ is the 
momentum of the scattered muon. It is instructive  
to carry out the calculation in the geometrical 
frame where the three-vector ${\bf V = p-k}$ is 
parallel to $z$-axis and the three-vector ${\bf k}$ is in the  $xz$-plane.  
We define the four-momentum  transfer to the nucleus as $q=P_i -P_f$. In
that frame the polar and axial angles of ${\bf q}$  
are denoted by  $\theta_q$ and $\phi_q$ respectively. 
 We use minus metric $\eta_{\mu \nu}=\mbox{diag}(+,-,-,-)$, in contrast to the authors of 
 Ref.~\cite{Liu:2017htz}, which implies that the  virtuality of the photon  is $t=-q^2=|{\bf q}|^2-q_0^2>0$. 
 After some algebraic manipulations one obtains the following expressions
\begin{equation}
\cos \theta_q = 
-\frac{|{\bf V}|^2+|{\bf q}|^2+m_{\mu}^2-(E_\mu+q_0-E_{Z'})^2}{2|{\bf V}| |{\bf q}|}, 
\end{equation} 
\begin{equation}
\qquad q_0 =-\frac{t}{2M}, 
\qquad |{\bf q}|= \sqrt{\frac{t^2}{4M^2}+t}.
\end{equation}
We assume that the nucleus has zero 
 spin~\cite{Liu:2017htz,Liu:2016mqv,Beranek:2013nqa,Beranek:2013yqa}, 
 as a result the photon-nucleus vertex is given by 
 \begin{equation}
i e F(t) (P_i+P_f)_\mu \equiv i e F(t) \P_\mu,
\end{equation}
where the squared elastic form-factor  is
\begin{equation}
F^2(t)  \equiv G^{el}_2(t)  \simeq Z^2 \left( \frac{a^2 t}{1+a^2t} \right)^2 
\left(\frac{1}{1+t/d}\right)^2,
\label{FFdefinition1}
\end{equation}
with $a=111 Z^{-1/3}/m_e$ and $d=0.164\, \mbox{GeV}^2 A^{-2/3}$. The contribution to the cross-section 
associated with squared inelastic form-factor is proportional to $Z$, therefore for heavy nucleus 
$Z\sim \mathcal{O}(100)$ that term is  negligible. In the case of NA64$\mu$, the active target is 
made of lead $(A=207, Z=82)$ being relevant only to the elastic form-factors giving the following 
typical momenta transfer associated with screening effects  and nucleus size respectively
\begin{equation}
    \sqrt{t_{a}}=1/a\simeq 2\cdot 10^{-5}\,\mbox{GeV},
    \label{tadefinition}
\end{equation}
\begin{equation}
    \sqrt{t_d}=\sqrt{d}\simeq 6.7\cdot 10^{-2}\,\mbox{GeV}.
    \label{tddefinition}
\end{equation}
We define the energy fraction  of the $Z'$-boson as $x=E_{Z'}/E_\mu$ 
and the angle between ${\bf k}$ and ${\bf p}$ as $\theta_{Z'}$.
One can express the differential cross-section 
in the following form~\cite{Liu:2017htz}
\begin{equation}
\label{eq:ETL_xtheta}
\frac{d\sigma}{dx \, d\cos \theta_{Z'}} =
 \frac{\epsilon^2  \alpha^3 |{\bf k}|E_\mu }{|{\bf p}|
|{\bf k - p}|  } \cdot  \int\limits_{t_{min}}^{t_{max}} \frac{d t}{t^2}
\, G_{2}^{el}(t) \cdot  \int\limits_{0}^{2\pi} \frac{d\phi_q}{2\pi}
\, \frac{|\mathcal{A}^{2\rightarrow 3}_{Z'}|^2}{8 M^2}, 
\end{equation}
where $\epsilon$ is  $Z'$ coupling to muons, which is related to $g'$ as follows  $\epsilon=g'/\sqrt{4\pi\alpha}$, here $\alpha=1/137$ is fine-structure constant, 
$t_{\text{min}}$ and $t_{\text{max}}$ are the values of minimum and 
maximum  momentum transfer respectively.
The quantities for 
$t_{\text{min}}$ and $t_{\text{max}}$
 are derived explicitly in Ref.~\cite{Liu:2017htz}. The
 production amplitude squared for vector dark boson such as $Z'$ is 
 calculated by using {\tt FeynCalc} tools~\cite{Shtabovenko:2020gxv,Shtabovenko:2016sxi} for 
 the {\tt Wolfram Mathematica} package~\cite{Mathematica} \begin{widetext}
\begin{equation}
\begin{gathered}
|\mathcal{A}^{2\rightarrow 3}_{Z'}|^2=
\frac{2}{\s^2 \u^2}
\Bigl(+\s \u \{\P^2 [(\s +t)^2+(\u+t)^2] -4 t[( \P \cdot p)^2 +( \P \cdot p')^2 ]\}\\
\\
+2 m_{\mu}^2 \{\P^2 t (\s+\u)^2-4[(\P \cdot p) \u +(\P \cdot p') \s]^2\}+
m_{Z'}^2 \{\P^2 t (\s-\u)^2-4[(\P \cdot p) \u +(\P \cdot p') \s]^2\}\Bigr),
\label{A2to3ZprETL}
\end{gathered}
\end{equation}
where the Mandelstam variables and relevant dot products are 
\begin{equation}
\s =(p'+ k)^2-m_{\mu}^2 = 2(p'\cdot k)+m_{Z'}^2, \qquad
\u =(p - k)^2-m_{\mu}^2 = - 2(p\cdot k)+m_{Z'}^2,
\end{equation}
\begin{equation}
\P^2 = 4 M^2+t, \qquad \P \cdot p = 2M E_\mu -(\s+t)/2, \qquad
\P \cdot p' = 2M(E_\mu-E_{Z'})+(\u-t)/2.
\end{equation}
\end{widetext}
The resulting amplitude squared (\ref{A2to3ZprETL}) coincides with the one given in 
Ref.~\cite{Liu:2017htz} for the case considered here, replacing the incident electron with a muon.

\section{The approximations for the
$Z'$ emission cross-sections
\label{WWCSSection}}
To calculate the differential $Z'$ emission cross-sections in the muon-nuclei interactions
we can use the so called Weiszäcker-Williams  approximation by assuming that the energy of
the initial particle is much higher than $m_{\mu}$ and $m_{Z'}$. In this case, the flux of
generated virtual photons can be considered as a plane wave and be approximated by a real
photon. This approximation allows thus to reduce the phase space of a $\mu(p)+N(\mathcal{P}_i) \to Z'(k)+\mu(p')+N(\mathcal{P}_f)$
process to a Compton-like~$\mu(p)+\gamma(q) \to Z'(k)+\mu(p')$ process~\cite{Kim:1973he, Gninenko:2017yus, Chen:2017awl}. In particular,
following the procedure described in Appendix D of~\cite{Kim:1973he}, which generalizes 
the classical WW one photon exchange process, we write down the following expression
\begin{widetext}
\begin{equation}
\frac{d\sigma(p+\mathcal{P}_i \to k +p' + \mathcal{P}_f )}{d(pk) d(k \mathcal{P}_i)}\Big|_{WW}   = 
  \frac{\alpha \chi }{\pi (p' \mathcal{P}_i)} \cdot \frac{d \sigma (p+q \to k+p')}{d(pk)} \Big|_{t=t_{min}}.
\label{WWTsai1}
\end{equation}
\end{widetext}
For ultra-relativistic incident muons we take into account that in the laboratory frame $d(k \mathcal{P}_i) = M d E_{Z'}$,  $(p' \mathcal{P}_i) = M  E_{\mu'}$,
$ d (pk) \simeq - |{\bf p}||{\bf k}| d \cos \theta_{Z'}$,  $|{\bf k}| = \sqrt{E_{Z'}^2 - m_{Z'}^2}$ 
and $E_{\mu'} \simeq E_\mu - E_{Z'}$. The WW approach  implies that 
the virtuality $t$ has its minimum  $t_{min}$ when $\q$ is collinear with 
$\k- \p$.
For the sake of simplicity we denote the cross-section on the left side of~Eq.(\ref{WWTsai1})
as $d \sigma_{2\to 3}$ and in the right side as $d \sigma_{2\to 2}$.
We obtain the following expression for the double 
differential cross-section: 
\begin{equation}
\frac{d\sigma_{2\to 3}}{dx \, d \cos \theta_{Z'}} \Big|_{WW} 
\simeq \frac{\alpha \chi }{\pi (1-x)} \cdot E_\mu^2 x \beta_{Z'}  \cdot  
 \frac{d \sigma_{2\to 2 }}{d(pk)} \Big|_{t=t_{min}},
 \label{BjorkensCorrected}
\end{equation}
where $\beta_{Z'} = \sqrt{1 - m_{Z'}^2/(x^2 E_\mu^2)}$. 
It is worth mentioning that the authors of Ref.~\cite{Bjorken:2009mm} made a typo in the 
$\beta_{Z'}$ definition. We take into account the $x$ dependence in $\beta_{Z'}$
as in Ref.~\cite{Liu:2017htz}, 
which plays an important role for $x\ll 1$.  
The effective photon flux $\chi$ in the WW  approach is defined by
\begin{equation}
\chi^{WW}= \int \limits_{t_{min}}^{t_{max}} d t \frac{t-t_{min}}{t^2} F^2(t),
\label{PhotonFlux1}
\end{equation}
where $ t_{min} \approx U^2(x,\theta_{Z'})/(4 E_\mu^2 (1-x)^2)$ and 
$t_{max} = m_{Z'}^2+m_{\mu}^2$ are the minimum and maximum squared momentum transfer 
to the nucleus. The expression for $t_{min}$ is derived below (for details, see e.~g.~Eq.~(\ref{AbsQdefinition}) and its discussion) and the function
$U(x,\theta_{Z'})$ is defined by  
\begin{widetext}
\begin{equation}
U \equiv 2(E_{Z'} E_\mu - |\k| |\p| \cos \theta_{Z'}) - m_{Z'}^2 \simeq
  E_\mu^2 \theta^2_{Z'} x + m_{Z'}^2 (1-x)/x + m_{\mu}^2 x. 
\label{uTildeDefForZpr}
\end{equation}
\end{widetext}
In Eq.~(\ref{uTildeDefForZpr}) we keep only leading terms in  
$m_{Z'}^2/E_{Z'}^2$, $m_{\mu}^2/E_{\mu'}^2$ and  $\theta_{Z'}^2$.
Considering the subsequent identities for the momentum squared
\begin{equation} M^2 = (\P_i -q)^2, \qquad
q^2 = (\P_i-\P_f)^2 = 2 M q_0, 
\end{equation}
implies that the expression for
the typical energy transferred to the nucleus can be expressed as:
\begin{equation}
 q_0\simeq - |\q|^2/(2 M).   
\label{ForQ0Definition}    
\end{equation}
This value is negligible and will be ignored in the calculation below. 
 One can easily obtain the  expression $ |\k-\p| \simeq E_\mu (1-x)$ by taking into account the 
 approximation discussed above. The next step is to consider 
the kinematic identity for the four-momentum squared of the outgoing muon: 
\begin{equation}
(p')^2 = (q-k+p)^2 \equiv m_\mu^2.    
\label{pSquaredOfOutMuon}
\end{equation}
 For $|\q|^2 \ll U$ and $|\q|^2 \ll  |\k-\p|^2$ Eq. (\ref{pSquaredOfOutMuon}) implies the relation between the 
 absolute value of nucleus momentum transfer and other kinematic variables
\begin{equation}
 |\q| \simeq U/(2 |\k-\p|) \simeq U/(2 E_\mu (1-x)).
 \label{AbsQdefinition}
 \end{equation}
 Note that in Eq.~(\ref{AbsQdefinition}) we take into account that  
the vectors $\q$ and $\k-\p$ are collinear. 
Therefore,  for WW approximation we get the
expression for the minimum nucleus momentum transfer, 
$t_{min}\simeq |\q|^2$.

The final ingredient to obtain the differential cross-section is the $\sigma_{2\rightarrow2}$ calculation. The differential cross-section in the Lorentz invariant notations~\cite{Bjorken:2009mm} is:
\begin{equation}
\frac{d \sigma_{2\to2}}{d(pk)} = \frac{d \sigma_{2\to 2}}{d (pp')} = \frac{2 \pi \alpha^2 \epsilon^2 }{\tilde{s}} \cdot |\mathcal{A}_{2\to2}^{Z'}|^2,     
\end{equation}
where~\cite{Liu:2017htz}
\begin{equation}
 |\mathcal{A}_{2\to2}^{Z'}|^2 = -2 \frac{\tilde{s}}{\tilde{u}}    -2 \frac{\tilde{u}}{\tilde{s}} 
 + 4(m_{Z'}^2 +2m_\mu^2)\left[  \left( \frac{\tilde{s}+\tilde{u}}{\tilde{s}\tilde{u}} \right)^2  m_\mu^2 
 - \frac{t_2 }{\tilde{s}\tilde{u}}\right]. 
 \label{AmplitudeElectronCase}
\end{equation}
where the Mandelstam variables are defined as follows  
\begin{equation}
\begin{gathered}
\u =      (p-k)^2 -m_\mu^2, \\ 
\\
\s = (p'+k)^2-m_\mu^2, \\  
\\
t_2 = (p-p')^2.  
\end{gathered}
\end{equation}
We note that authors of Ref.~\cite{Bjorken:2009mm}
dropped  terms in~(\ref{AmplitudeElectronCase}) which are associated with a squared
incident particle mass (the electron mass term in Ref.~\cite{Bjorken:2009mm}). As 
explained  in~\cite{Liu:2017htz}, dropping these terms will lead to an underestimation of the experiment sensitivity region. Moreover, when the incident particle is a muon, one can not drop the 
corresponding  terms~\cite{Liu:2017htz} since the underestimation of the available parameter space will be significant as soon as $m_{Z'} \lesssim m_{\mu}$. 
Finally by using the relation for the scalar product $(qk) \simeq U x/(2 (1-x))$ and the approximate expression 
$\s+t_2+\u \simeq  m_{Z'}^2$ one can easily get 
\begin{equation}
\u \simeq   - U, \quad \s \simeq U/(1-x), \qquad t_2 \simeq - x U/(1-x) +m_{Z'}^2.  
\label{MandelstamResulted1}
\end{equation}
We use Eqs.~(\ref{MandelstamResulted1}) below in Sections~\ref{labelWWsubsection} 
and~\ref{labelIWWsubsection}  
to derive the double differential cross-section in both WW and IWW approach. 


\subsection{Cross-section in WW approximation
\label{labelWWsubsection}}
The photon flux can be integrated numerically according to Eq.~(\ref{PhotonFlux1}). 
However, it is instructive and less computational demanding to develop an analytical 
expression for this equation by using the specific form of elastic form-factor, Eq. (\ref{FFdefinition1}). 
In this case, one can  get the following analytical expression for the effective photon flux
\begin{equation}
\chi = Z^2[\tilde{\chi}(t_{max})-\tilde{\chi}(t_{min})],
\label{exact_Chi_1}
\end{equation}
 where 
 \begin{widetext}
 \begin{equation}
\tilde{\chi}(t)=\frac{t_d^2 }{(t_a-t_d)^3}\Bigl[\frac{(t_a-t_d) (t_a+t_{min})}{t+t_a}+\frac{(t_a-t_d)
   (t_d+t_{min})}{t+t_d}
   +(t_a+t_d+2
   t_{min}) \log\left( \frac{t+t_d}{t+t_a}\right) \Bigr],
 \end{equation}
 \end{widetext}
which is valid  for both WW and IWW approaches. Here $t_a$ and $t_d$ are defined by Eq.~(\ref{tadefinition}) and 
Eq.~(\ref{tddefinition}) respectively. 
 However, in this subsection we specify the photon flux for the case of double differential WW cross-section,
that simply implies that $ t_{min} \approx U^2(x,\theta_{Z'})/(4 E_\mu^2 (1-x)^2)$. 
In particular,  one has the following expression for the WW cross-section of $Z'$ boson
\begin{widetext}
\begin{equation}
\left(\frac{d \sigma_{2\to 3}}{dx \, d\cos \theta_{Z'}} \right)_{WW}=
2 \epsilon^2 \alpha^3 \sqrt{x^2-\frac{m_{Z'}^2}{E_\mu^2}} E^2_\mu (1-x) \frac{\chi^{WW}}{\tilde{u}^2} 
|\mathcal{A}^{Z'}_{2\to 2}|^2.
\label{WWZprFinalNumerical}
\end{equation}
By substituting Eqs.~(\ref{uTildeDefForZpr}) and (\ref{MandelstamResulted1}) into 
Eq.~(\ref{AmplitudeElectronCase}), the amplitude squared can be rewritten as~\cite{Liu:2017htz}
\begin{equation}
\label{ListOfAmpSq}
|\mathcal{A}^{Z'}_{2\to 2}|^2 = 2\frac{2-2x+x^2}{1-x}+4(m_{Z'}^2+2m_\mu^2)\frac{\tilde{u}x+m_{Z'}^2(1-x)+m_\mu^2x^2}{\tilde{u}^2}.
\end{equation}
\end{widetext}
Now let us specify the phase space for the 
produced $Z'$. In particular, Eq.~(\ref{WWZprFinalNumerical}) implies that $x_{min} \equiv m_{Z'}/E_{\mu} \lesssim x$.  The nuclear 
transfer energy is given by Eq.~(\ref{ForQ0Definition}), so that it can be neglected for 
the   large $Z'$ energy fraction bound $x\lesssim 1$. 
 Therefore one can immediately get the following upper bound on  $x  \lesssim x_{max} \equiv 1-m_{\mu}/E_\mu$.  
\subsection{Calculations using IWW approximation
\label{labelIWWsubsection}}
In the so-called improved WW approach the dependence of $t_{min}$ on $x$ 
 and $\theta_{Z'}$ in the flux derivation (Eq.~(\ref{PhotonFlux1})) is omitted to simplify calculations, which is important in Monte-Carlo (MC) simulations, such that
 $t^{IWW}_{min} \simeq m_{Z'}^4/(4 E_{\mu}^2)$. 
 This means however that the IWW approach is less accurate~\cite{Liu:2017htz}. 

\begin{figure*}[]
    \centering
    \includegraphics[width=0.45\textwidth]{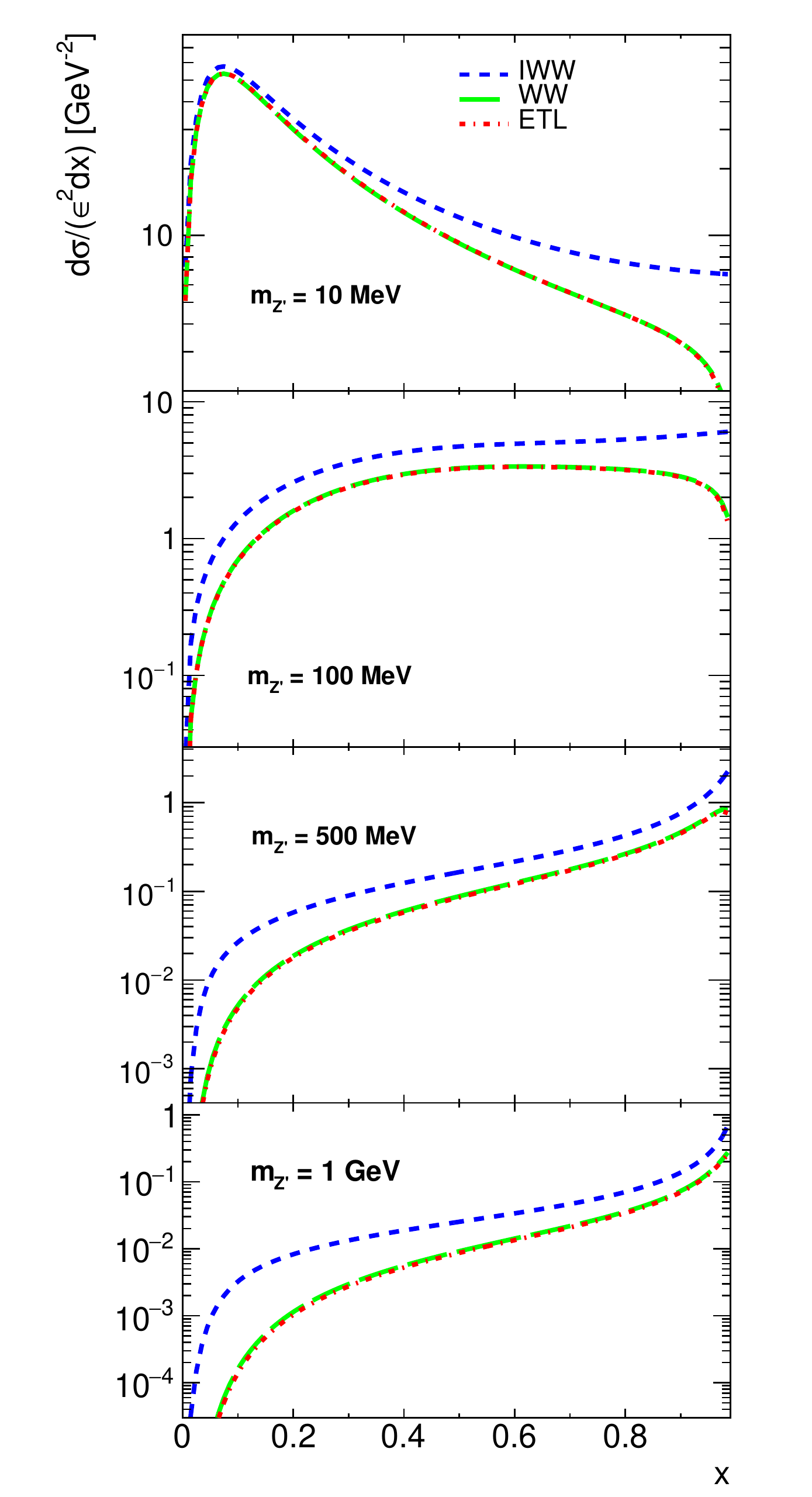}
    \hspace{5mm}
    \includegraphics[width=0.45\textwidth]{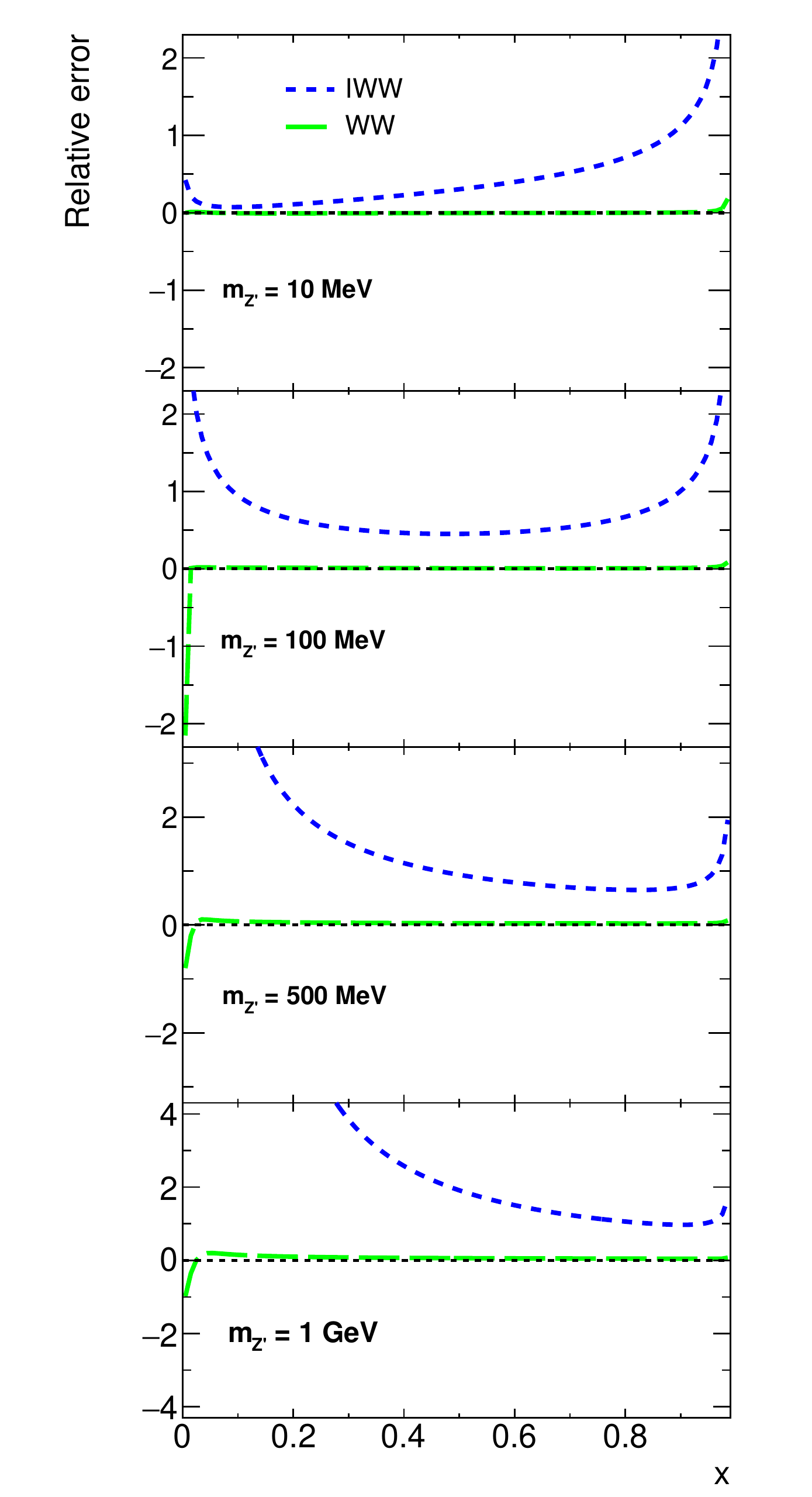}
    \caption{({\it Left}) Differential cross-section as a function of $x$ for IWW (blue dashed line), WW (green line) and ETL (red dotted line) for different $Z'$ masses. The single- and double-differential cross-sections (see e.g. Eqs.~(\ref{eq:IWW_x}), (\ref{eq:WW_x}) and (\ref{eq:ETL_xtheta})) are numerically integrated using respectively adaptive and Monte-Carlo procedures \cite{Galassi}.  ({\it Right}) Relative error between WW and IWW approximations as a function of $x$ defined as  $(\mathcal{O}_{approx}-\mathcal{O}_{exact})/\mathcal{O}_{exact}$ (similar approach than the one used in \cite{Liu:2017htz}).
    \label{fig:dsdx_Z}}
\end{figure*}

We compute the differential cross-section using the IWW approximation
to compare with the approach developed in the previous section. The 
integration over $\theta_{Z'}$ in this case can be performed analytically 
(see, e.g. Eq.~(30) in Ref.~\cite{Liu:2017htz}). 
In particular, one has
\begin{widetext}
\begin{equation}
    \label{eq:IWW_x}
    \left(\frac{d\sigma_{2\to 3}}{dx}\right)_{IWW}=2\epsilon^2 \alpha^3\chi^{IWW}
    \sqrt{x^2-\frac{m_{Z'}^2}{E_\mu^2}}\left[ \frac{m_\mu^2x(-2+2x+x^2)-2(3-3x+x^2)\tilde{u}}{3x\tilde{u}^2} \right] \Bigr|^{\tilde{u}_{max}}_{\tilde{u}_{min}}
\end{equation}
\end{widetext}
 where $\tilde{u}_{max}=-m_{Z'}^2\frac{1-x}{x}-m_\mu^2 x$ and 
 $\tilde{u}_{min}=-x E_{\mu}^2 (\theta^{max}_{Z'})^2-m_{Z'}^2\frac{1-x}{x}-m_\mu^2 x$.
 In contrast to Ref.~\cite{Liu:2017htz}, we keep 
 $\tilde{u}_{min}$ finite to take into account the effects associated with the maximal angle of emitted $Z'$.  As discussed before, in the IWW approximation 
$\chi$ is independent
of $x$ and  $\theta_{Z'}$. The resulting IWW cross-section has a sharp peak at $x\simeq 1$, as illustrated in Fig. \ref{fig:dsdx_Z}. The numerical calculations reveal that this peak is smeared in WW approach due to the fact that $\chi$ is a function of $x$ and  $\theta_{Z'}$ (see the green line in the right plot of Fig. \ref{fig:dsdx_Z}).

\begin{figure*}[]
    \centering
    \includegraphics[width=0.46\textwidth]{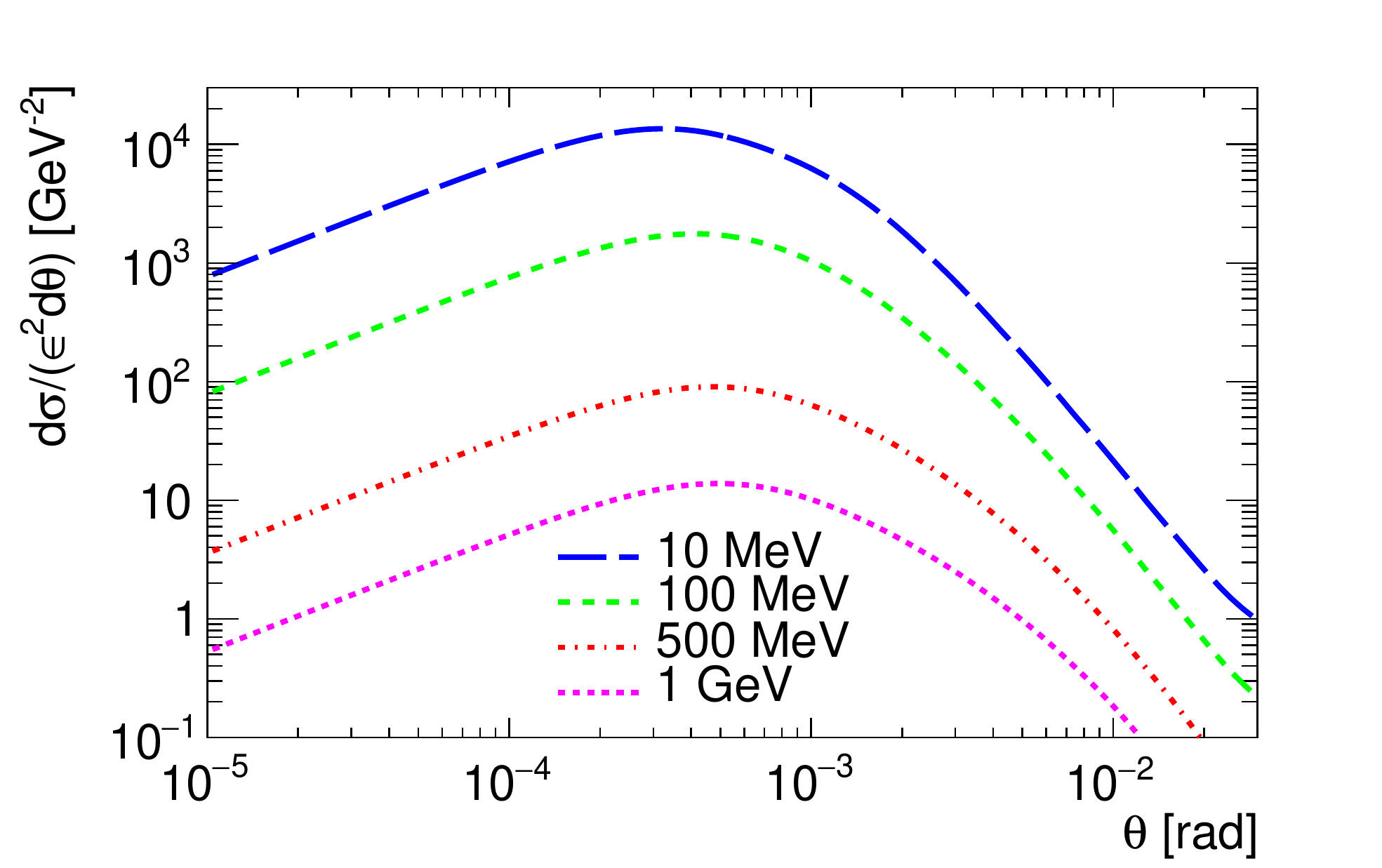}
\hspace{0.5mm}
\includegraphics[width=0.46\textwidth]{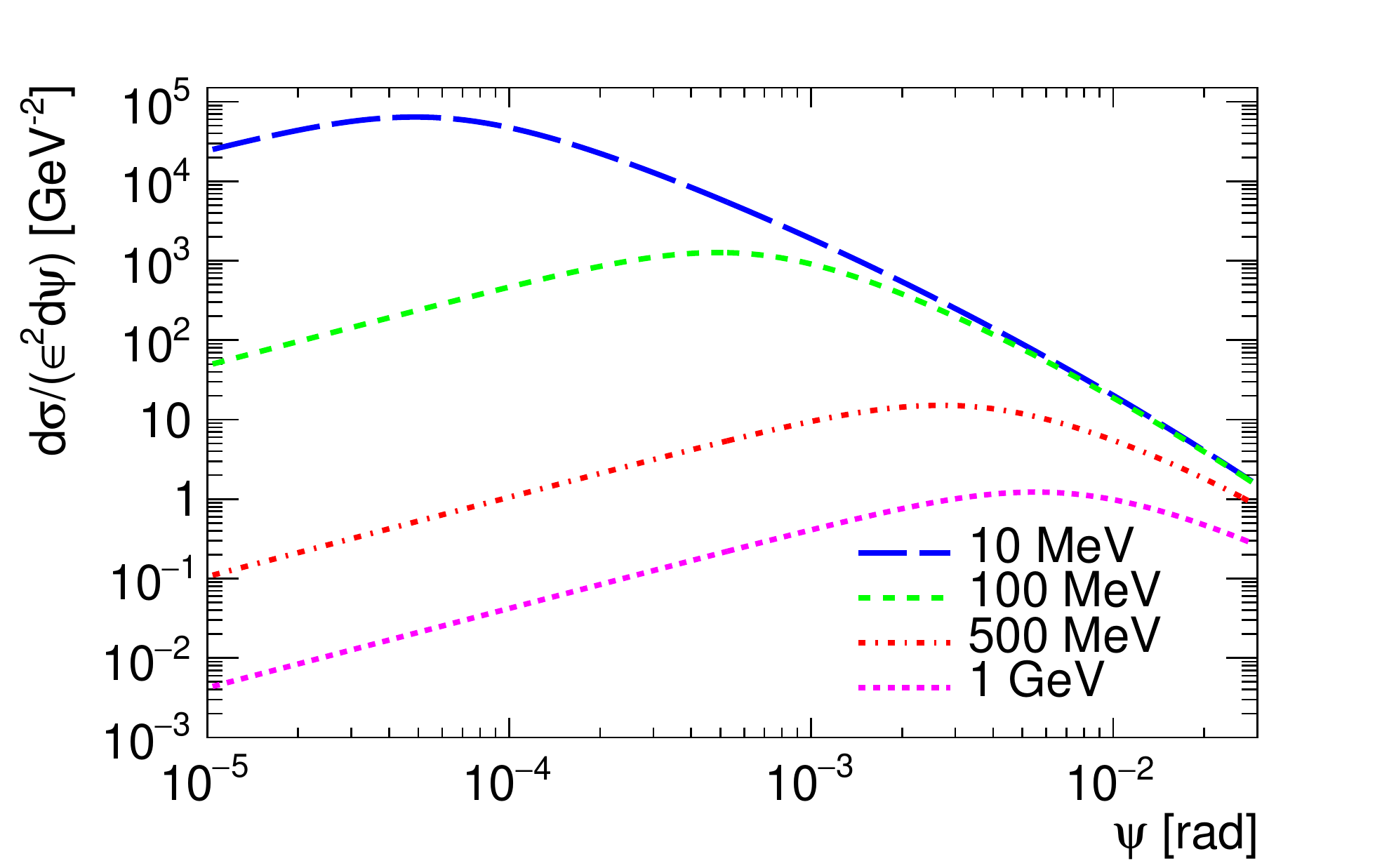}
\caption{Differential cross-sections in the WW approach as a function of  $\theta$ ({\it Left}) and $\psi$ ({\it Right}), peaking respectively around $\theta_{Z'}\simeq m_\mu/E_\mu$ and $\psi_{\mu'}\simeq m_{Z'}/E_\mu$. 
\label{fig:angular_spectra}
}
\end{figure*}

 The comparison between the differential cross-section as a function of 
 $x$ calculated using the three approaches for different $Z'$ masses is 
 shown in Fig. \ref{fig:dsdx_Z}.
 The muon energy has been set to 160 GeV as expected in NA64$\mu$ experiment. We note that the typical angle of $Z'$ emission does not depend on its mass as it scales as 
 $\theta_{Z'}\sim m_{\mu}/E_{\mu}$. This can be seen in Fig. \ref{fig:angular_spectra}, left plot. Thus, the differential 
 cross-sections (see e.g. Eqs.~(\ref{eq:IWW_x}), (\ref{eq:WW_x}) and 
 (\ref{eq:ETL_xtheta})) are numerically integrated using respectively 
 adaptive and Monte-Carlo procedures \cite{Galassi} up to a maximum $Z'$ 
 angle $\theta^{max}_{Z'}\sim0.1$. 
  
 The values obtained for the differential cross-section are different depending on 
 the mass because of the dynamic conditions. In the right plots of Fig. \ref{fig:dsdx_Z}, the 
 relative difference between the approximations with respect to the the 
 exact-tree-level calculation is shown. The differences between WW and the exact calculation
 are below 2$\%$ in the mass region between $10$~MeV and $1$~GeV. 
 Deviations from exact cross-sections are significant in the case of IWW 
 approximation, especially in the extreme fractional energy regions, due to the flux simplification. 

\begin{figure*}[]
    \centering
    \includegraphics[width=0.45\textwidth]{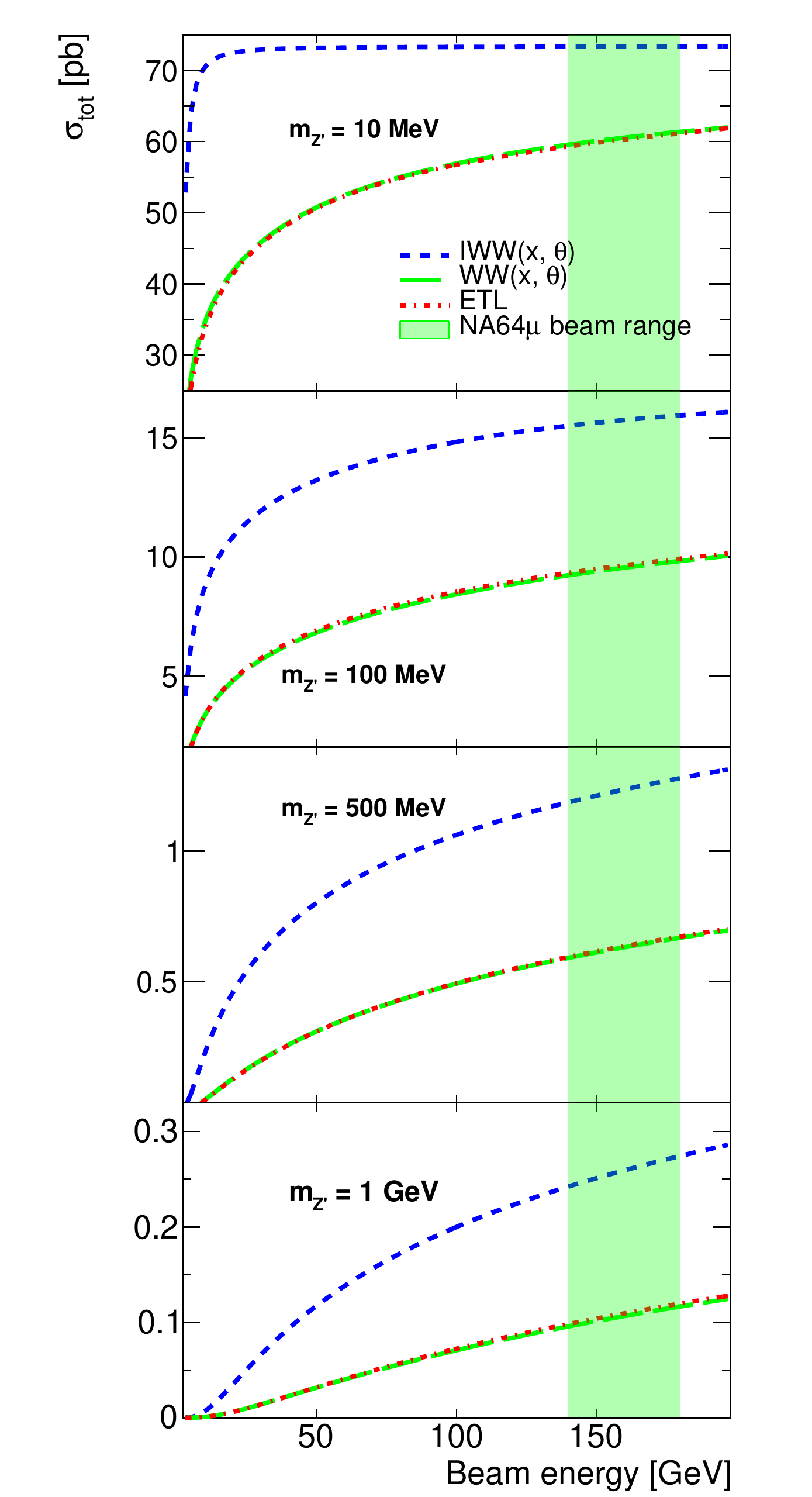}
    \hspace{5mm}
   \includegraphics[width=0.45\textwidth]{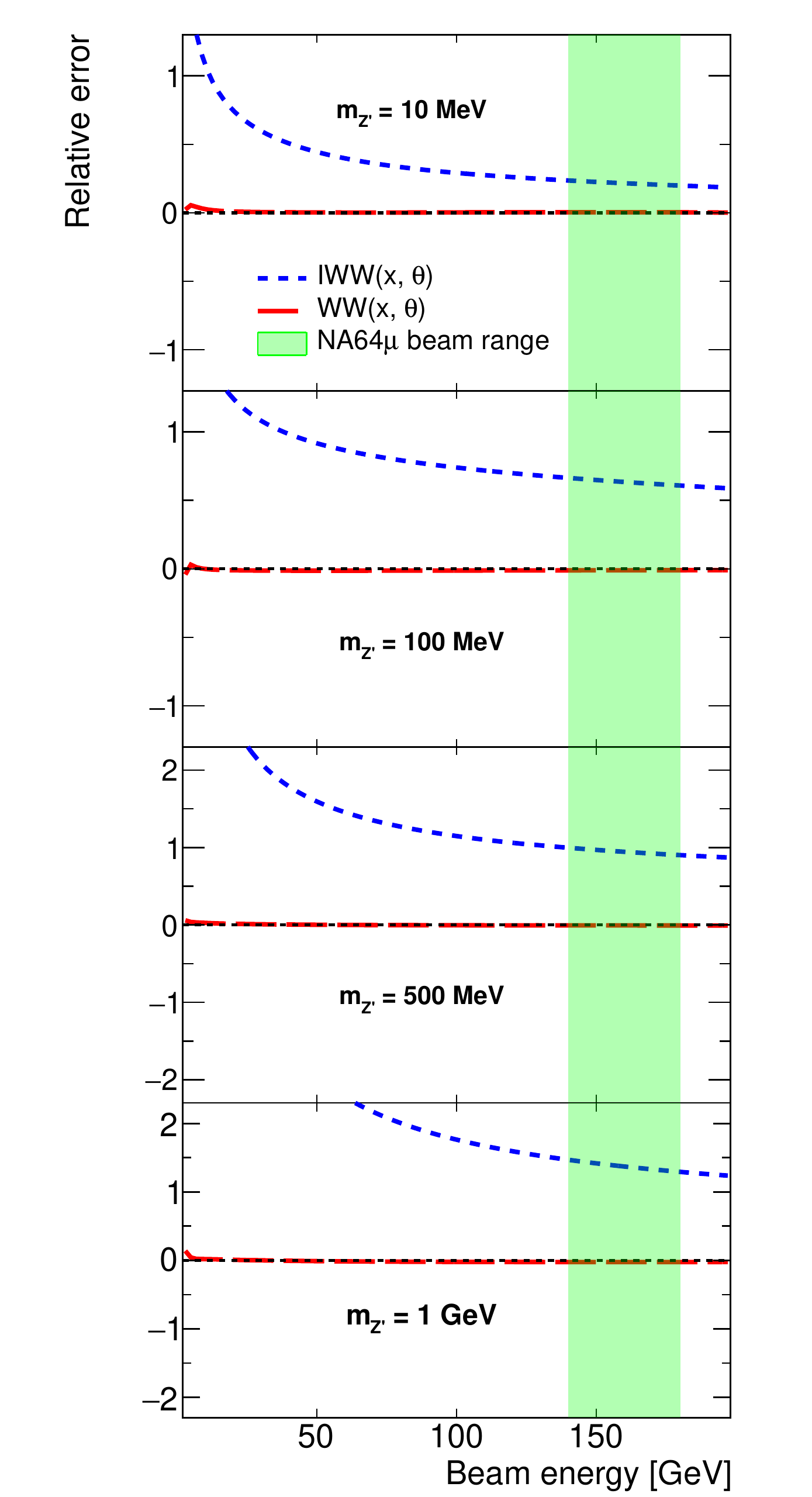}
    \caption{({\it Left}) Total cross-section as a function of the beam energy for IWW (blue dashed line), WW (green line) expressions and ETL (red dotted line) integrated over $x,\ \theta$.  ({\it Right}) Relative error between WW, IWW approximations and ETL as a function of the beam energy expressed as  $(\mathcal{O}_{approx}-\mathcal{O}_{exact})/\mathcal{O}_{exact}$. The typical NA64$\mu$ beam energy range is shown as a green band. 
    \label{fig:totX_Z}}
\end{figure*}

 In Fig. \ref{fig:totX_Z} the evolution of the total cross-section, calculated with the different approaches, is presented as a function of the initial muon energy for different $Z'$ masses. The WW approximation reproduces the ETL cross-section with a relative difference below 2$\%$ for the whole mass range. For illustration, the energy region interesting for the NA64$\mu$ experiment is highlighted in green. In this region, the relative differences between IWW and ETL (Sec. \ref{ETLSection}) are smaller than in the low energy region, where they tend to diverge, when the muon energy approaches the $Z'$ mass.

$$
$$
\subsection{Zero mass limit for WW approach}
An independent cross-check to validate our calculations, is to consider the case of massless $Z'$,  $(m_{Z'}\rightarrow 0)$, since this should reproduce the muon nuclear bremsstrahlung $\mu N \to \mu N \gamma$ cross-section~\cite{Groom:2001kq}. We have used a simplified analytical expression of the flux for these calculations valid for the parameter space of interest $t_{max}\gg t_d \gg t_{min}$ and $t_d \gg t_a$ with logarithmic accuracy:

\begin{equation}
    \frac{\chi}{Z^2} \simeq \log\left(\frac{t_d}{t_{min}+t_a}\right)-2.
    \label{chiApprox2}
\end{equation}

We note that this expression is no longer valid as $t_{min} \simeq  t_{d}$, which implies large masses
$\mathcal{O}(1)$~GeV, low energies $E_\mu \lesssim 10$~GeV and $x\simeq 1$. However, for the interesting NA64$\mu$ parameter space, one can use the approximated formula for the calculation of the analytical differential cross-section $(d\sigma/dx)_{WW}$. 

\begin{figure*}[]
\centering
\includegraphics[width=0.46\textwidth]{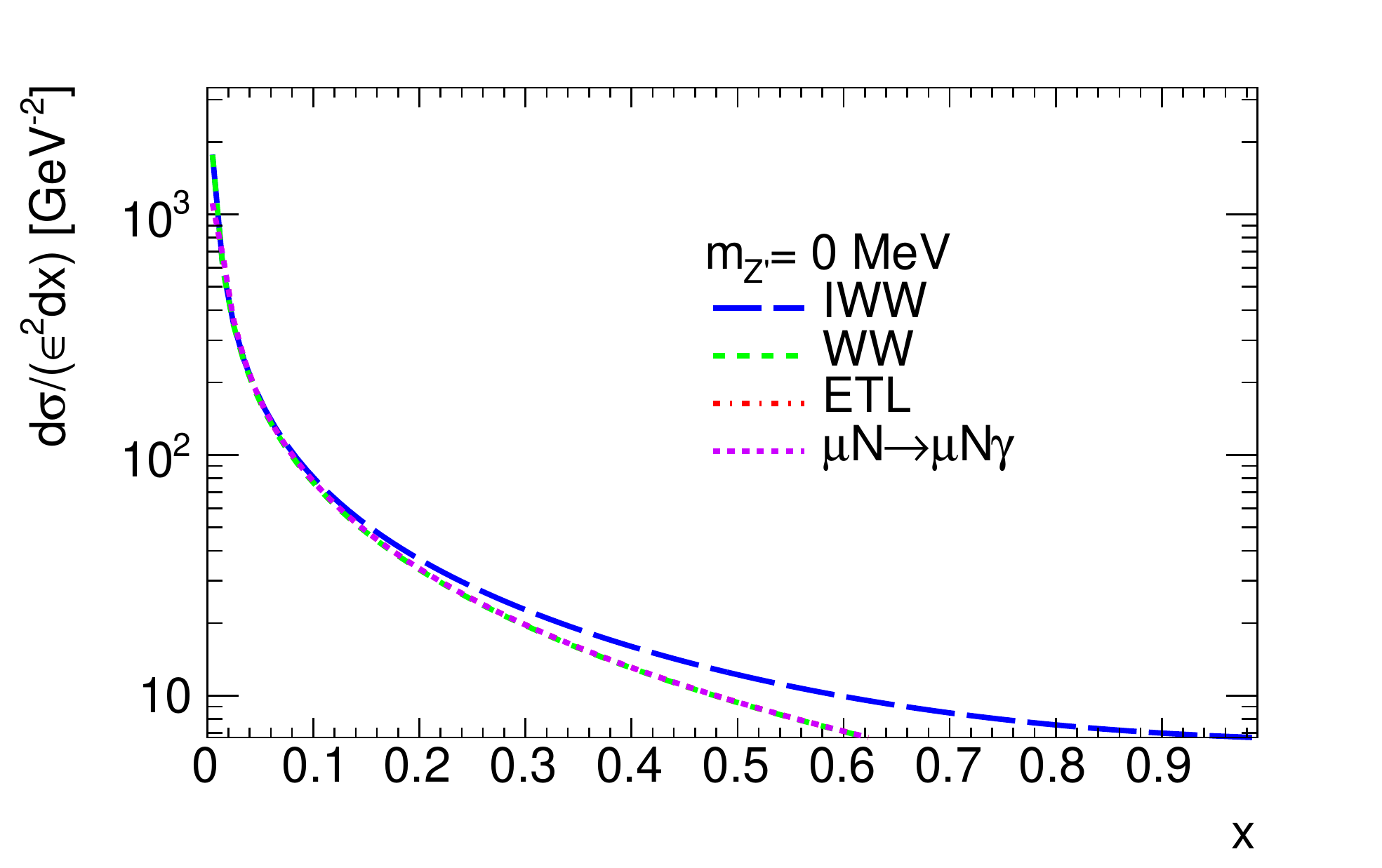}
\hspace{0.5mm}
\includegraphics[width=0.46\textwidth]{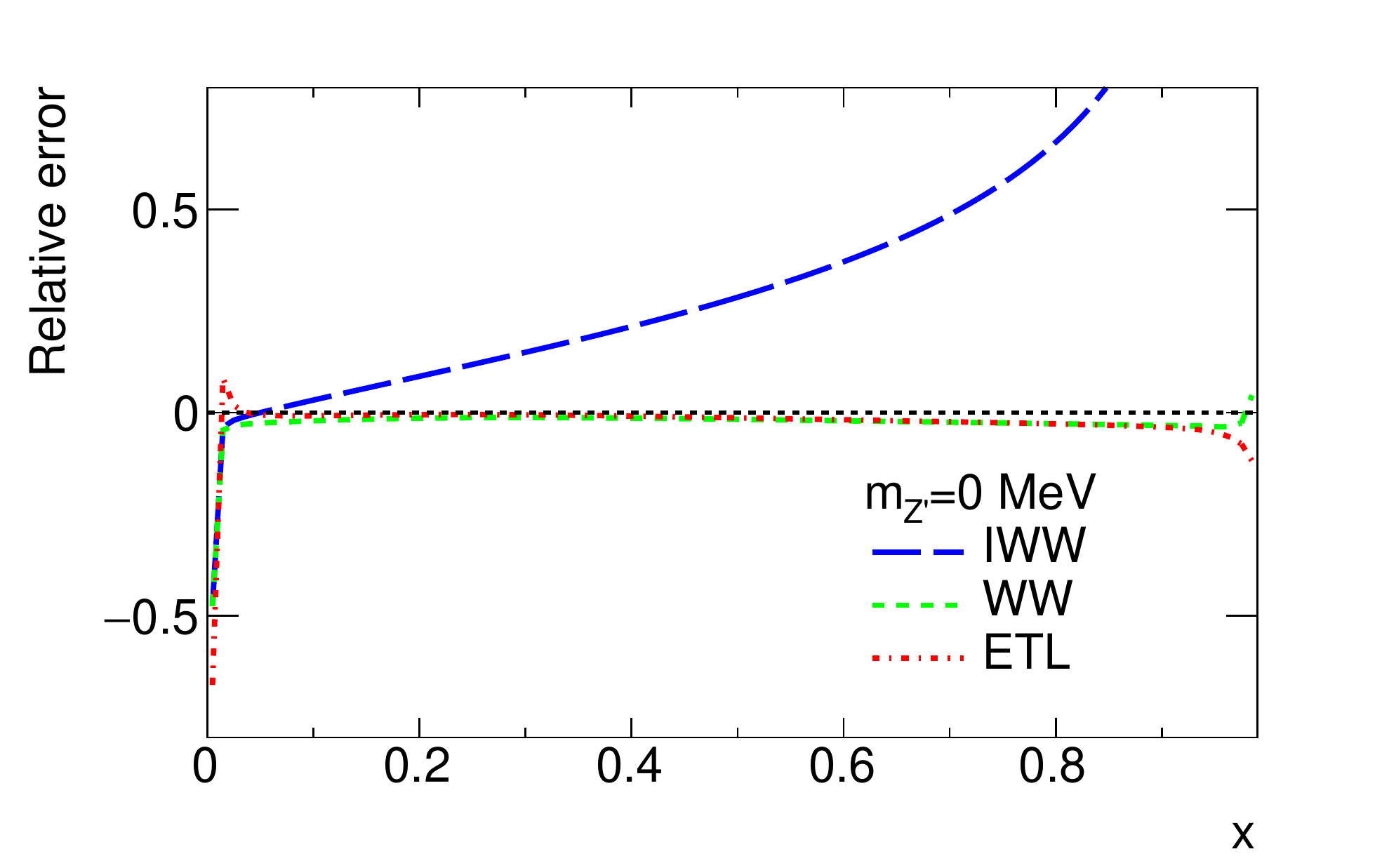}
\caption{({\it Left}) Differential cross-sections as a function
of $Z'$ energy fraction $x$ calculated at ETL and in WW and IWW approaches for $m_{Z'}=0$ and $\epsilon=1$. Muon 
nuclear bremsstrahlung $\mu N \to \mu N \gamma$ 
cross-section~\cite{Kelner95} is shown as a purple dotted line. One
can see that as soon as $m_{Z'}\to 0$ the shape of 
the muon bremsstrahlung $\mu N \to \mu N \gamma$ is
reproduced by $\mu N \to \mu N Z'$. ({\it Right}) Relative error defined as $(\mathcal{O}_{m=0}-\mathcal{O}_{brem.})/\mathcal{O}_{brem.}$.
The angle of $Z'$ emission is integrated in the range $\theta_{Z'}\in[0,0.1]$. 
\label{fig:dsdx_ZandMuonBrems_ETL}
}
\end{figure*}

In particular, the numerical calculation reveals that for $m_{A'} \lesssim 500$~MeV and $E_{\mu} \simeq 150-160$~GeV the discrepancy between the
cross-sections calculated with (\ref{chiApprox2}) and (\ref{exact_Chi_1}) is well below $1 \%$. We derive the approximate analytical formula for $Z'$ production cross-section using WW approximation 
for this particular case considering only an elastic nuclear form-factor. The first step is to integrate out $\theta_{Z'}$ in~Eq.~(\ref{ListOfAmpSq}). We obtain 
\begin{widetext}
\begin{equation}
    \label{eq:WW_x}
    \left(\frac{d \sigma}{dx} \right)_{WW} = \epsilon^2 \alpha^3 \sqrt{x^2 -m_{Z'}^2/E_\mu^2} \,\, \frac{(1-x)}{x} \int\limits^{\tilde{u}_{max}}_{\tilde{u}_{min}}
    d \tilde{u} \frac{|\mathcal{A}^{Z'}_{2\to 2}|^2}{\tilde{u}^2} \chi^{WW}.
\end{equation}
 Finally, we get
 \begin{equation}
      \left(\frac{d \sigma}{dx} \right)_{WW} = \epsilon^2 \alpha^3 Z^2 \sqrt{x^2 -m_{Z'}^2/E_\mu^2} \,\, \frac{(1-x)}{x}
      \times   \left\{ 
      2\frac{2-2x+x^2}{1-x} I_2+4(m_{Z'}^2+2m_\mu^2) \left[x I_3+  (m_{Z'}^2(1-x)+m_\mu^2x^2)I_4\right]
      \right\}.
 \end{equation}
where the auxiliary integrals are 
\begin{equation}
 I_n =  \int\limits^{\tilde{u}_{max}}_{\tilde{u}_{min}}
    d \tilde{u}\, \frac{1}{\tilde{u}^n} \log \left[ \frac{\beta_d^2 }{\tilde{u}^2 + \beta_a^2} \right], \quad \mbox{for} \quad n=2,3,4.
\end{equation}
In this expression $\beta_d^2 = 4 E_{\mu}^2 (1-x)^2 t_d/e^2$, being $e\simeq 2.71828$ the Euler's 
number, $\beta_a^2=4 E_{\mu}^2 (1-x)^2 t_a$ and 
$\tilde{u}_{min}=-x (\theta^{max}_{Z'})^2 E_\mu^2-m_{Z'}^2\frac{1-x}{x}-m_\mu^2 x$. The resulting integrals  for different $n$ values are:
 \begin{equation}
I_2 =\left[ -\frac{1}{\tilde{u}}\log \left(\frac{\beta_d^2}{\beta _a^2+\tilde{u}^2}\right)-\frac{2}{\beta_a} \arctan\left(\frac{\tilde{u}}{\beta_a}\right) \right] \Bigg|^{\tilde{u}_{max}}_{\tilde{u}_{min}},
 \end{equation}
  \begin{equation}
I_3=  \left[-\frac{1}{2 \tilde{u}^2} \log
   \left(\frac{\beta_d^2}{\beta_a^2+\tilde{u}^2}\right)  + \frac{1}{2 \beta_a^2} \log \left(\frac{\beta_a^2+\tilde{u}^2}{\tilde{u}^2}\right) \right]\Bigg|^{\tilde{u}_{max}}_{\tilde{u}_{min}},  \end{equation}
 \begin{equation}
    I_4 =\left[-\frac{1}{3
   \tilde{u}^3} \log \left(\frac{\beta_d^2}{\beta_a^2+\tilde{u}^2}\right)+\frac{2}{3 \tilde{u} \beta_a^2}+\frac{2}{3 \beta_a^3}  \arctan\left(\frac{\tilde{u}}{\beta_a}\right) \right] \Bigg|^{\tilde{u}_{max}}_{\tilde{u}_{min}}.
 \end{equation}
\end{widetext}
In Fig. \ref{fig:dsdx_ZandMuonBrems_ETL} the differential cross-sections for the three approaches in the 
low $Z'$ mass region are compared to the cross section obtained for the QED muon bremsstrahlung process 
(purple dotted line). The shape of the distributions is identical, and the relative differences of the WW and ETL distributions with respect to the muon nuclear bremsstrahlung are below~2$\%$. 


\section{Differential cross-section for the scattered muon
\label{MuonWWCSSection}}

For the successful detection of $Z'$ in fixed target experiments such as NA64$\mu$, it is 
crucial to tag the scattered muon after $Z'$ emission. In this section, we focus on calculating the differential cross section as a function of the scattered muon  angle $\psi_{\mu'}$ to evaluate its possible impact in the expected signal yield. In the WW approximation, the differential cross-section for the process 
$\mu(p)+N(\mathcal{P}_i) \to Z'(k)+\mu(p')+N(\mathcal{P}_f)$ is:
\begin{equation}
\frac{d \sigma }{d \cos \psi_{\mu'} d y} = \frac{\alpha \chi }{\pi}\frac{E_\mu^2 y \beta_{\mu '}}{(1-y)}  \times 
  \frac{d \sigma_{2\to2} }{d (p p')} \Big|_{t=t_{min}},
  \label{DsDyDpsi1}
\end{equation}
where 
$y = E_\mu'/E_\mu$ and $\beta_{\mu '} = \sqrt{1-m_\mu^2/(E_{\mu}^{'2})}$ are
the fraction energy and the velocity of the outgoing muon respectively. 
One can obtain the following differential cross-section of the Compton-like 
process $e \gamma \to e Z'$, which is written in Lorentz-invariant form
\begin{widetext}
\begin{equation}
 \frac{d \sigma_{2\to 2} }{d (p p')} \Big|_{t=t_{min}} = 
 \frac{4 \pi \alpha^2 \epsilon^2}{\tilde{s}^2}\times 
 \left(   - \frac{\tilde{s}}{\tilde{u}}    - \frac{\tilde{u}}{\tilde{s}} 
 + 2(m_{Z'}^2 +2m_\mu^2)\left[  \left( \frac{\tilde{s}+\tilde{u}}{\tilde{s}\tilde{u}} \right)^2  m_\mu^2 
 - \frac{t_2 }{\tilde{s}\tilde{u}}\right]  \right).
 \label{dydpsiWWmuon1}
\end{equation}
\end{widetext}
The WW approach implies that the cross-section (\ref{dydpsiWWmuon1}) is calculated 
for the minimum value of the virtuality $t_{min}$, i.~e.~when $\q$ is 
collinear with  $\p'- \p$. The later condition in association with the relation 
$k^2 \equiv m_{Z'}^2 = (q-(p' -p))^2$ yields
\begin{equation}
 2 |\q | |\p' - \p|   +t_2 \simeq 
m_{Z'}^2.     
\end{equation}
One can write down the relation $ |\p'-\p| \simeq E_\mu (1-y)$,
which is valid as soon as $m_{\mu}\ll E_{\mu}, E_{\mu'}$ and $\psi_{\mu'}\ll 1$.
Therefore it yields 
\begin{equation}
t_{min}= |\q|^2, \qquad |\q| = \frac{m_{Z'}^2 - t_2}{2E_\mu (1-y)}.
\label{tminWW}
\end{equation}
We also use below the following notation for the auxiliary Mandelstam variable $\tilde{t}=m_{Z'}^2 -t_2$.
The Mandelstam variables $\tilde{s}$, $\tilde{u}$ and $t_2$ can be written then as
\begin{widetext}
\begin{equation}
\begin{gathered}
\tilde{s}= (p'+k)^2-m_\mu^2 =\tilde{t}/(1-y),\\
\\
\u =(p' - q)^2 -m_\mu^2 \simeq -2 (p' q) \simeq 
2 |\p'| |\q| \cos \theta_{p'q} \simeq - 2 |\p'| |\q|   
\simeq - y\, \tilde{t}/(1-y),\\
\\
t_2 = - [ E_\mu^2 \psi^2_{\mu'} y + m_\mu^2 (1-y)/y + m_\mu^2 y ] +m_\mu^2.
\end{gathered}
\end{equation}
\end{widetext}
In the WW approach the effective photon flux $\chi$ in~(\ref{DsDyDpsi1}) is defined 
 in~(\ref{PhotonFlux1}) with $t_{min}$ to be calculated using~(\ref{tminWW}) and $t_{max}=m_{Z'}^2+m_\mu^2$.   
Finally, the differential cross-section as a function of the muon fractional energy $y$ and the recoil muon angle $\psi_{\mu'}$ (Eq.~(\ref{DsDyDpsi1}), can be written as 
follows
\begin{widetext}
\begin{equation}
       \frac{d \sigma }{d \cos \psi_{\mu'} d y} = 
    8 \alpha^3 \epsilon^2 \chi^{WW} E_\mu^2 \sqrt{y^2-m_\mu^2/E_{\mu}^2} \cdot 
    \frac{(1-y)}{\t^2}\cdot  \left[
    \frac{1}{2y} + \frac{y}{2} + \left( m_{Z'}^2 +2 m_{\mu}^2  \right) \frac{(1-y)^2}{\t^2 y } \left( m_{\mu}^2 \frac{(1-y)^2}{y} +m_{Z'}^2 - \t \right) 
    \right]. 
      \label{DsDyDpsi2}
\end{equation}
\end{widetext}


The next step is to define our integration limits. One can see from  Eq.~(\ref{DsDyDpsi2}) that the minimal muon fractional energy after emission is $y_{min} \simeq m_\mu/E_\mu$, when almost all energy of the initial muon is transferred to $Z'$. On the other hand, when the muon transfers close to zero energy to $Z'$ we have $y_{max} \simeq 1- m_{Z'}/E_\mu$. 

\begin{figure*}[]
    \centering
    \includegraphics[width=0.45\textwidth]{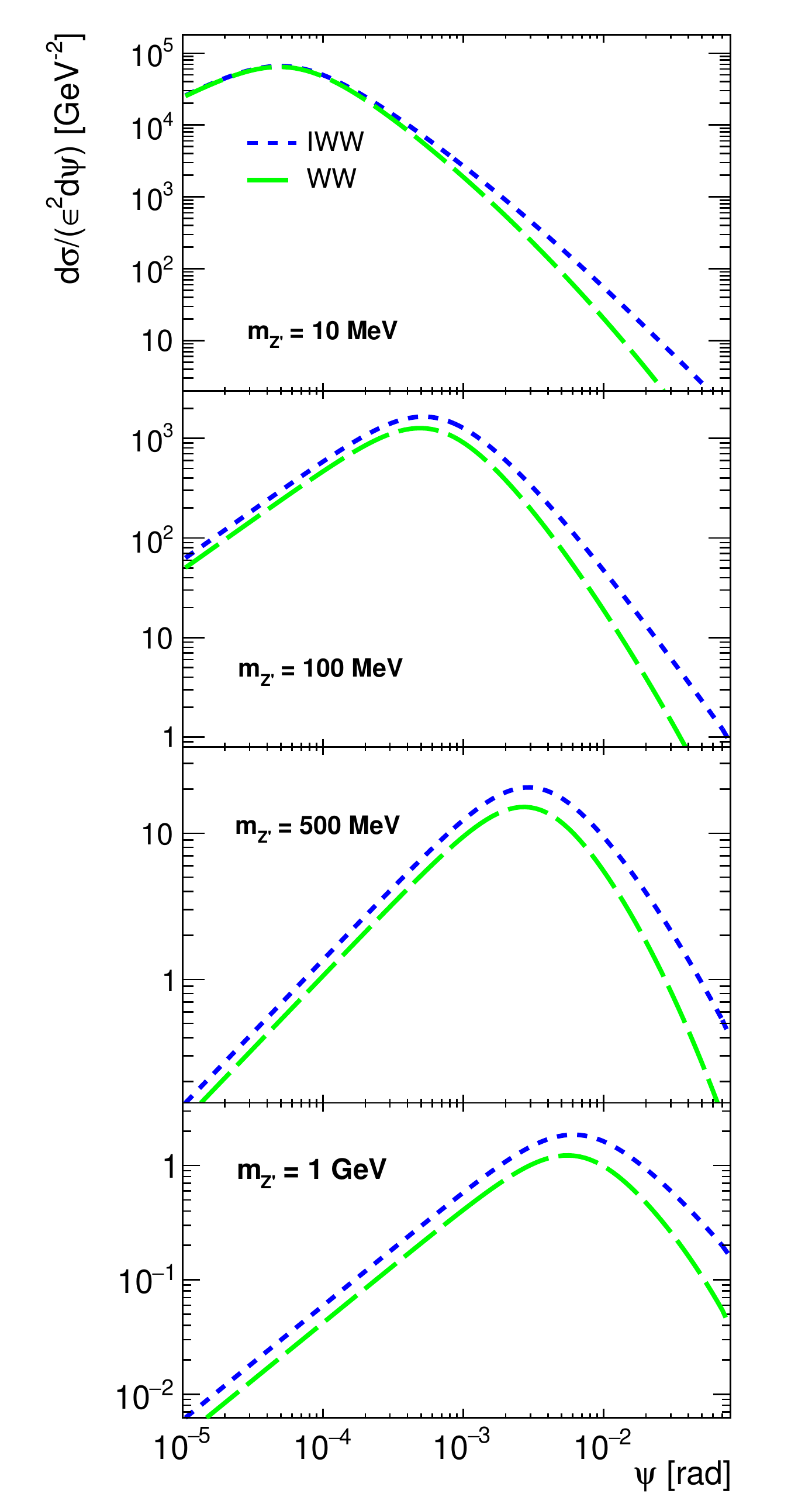}
    \hspace{5mm}
    \includegraphics[width=0.45\textwidth]{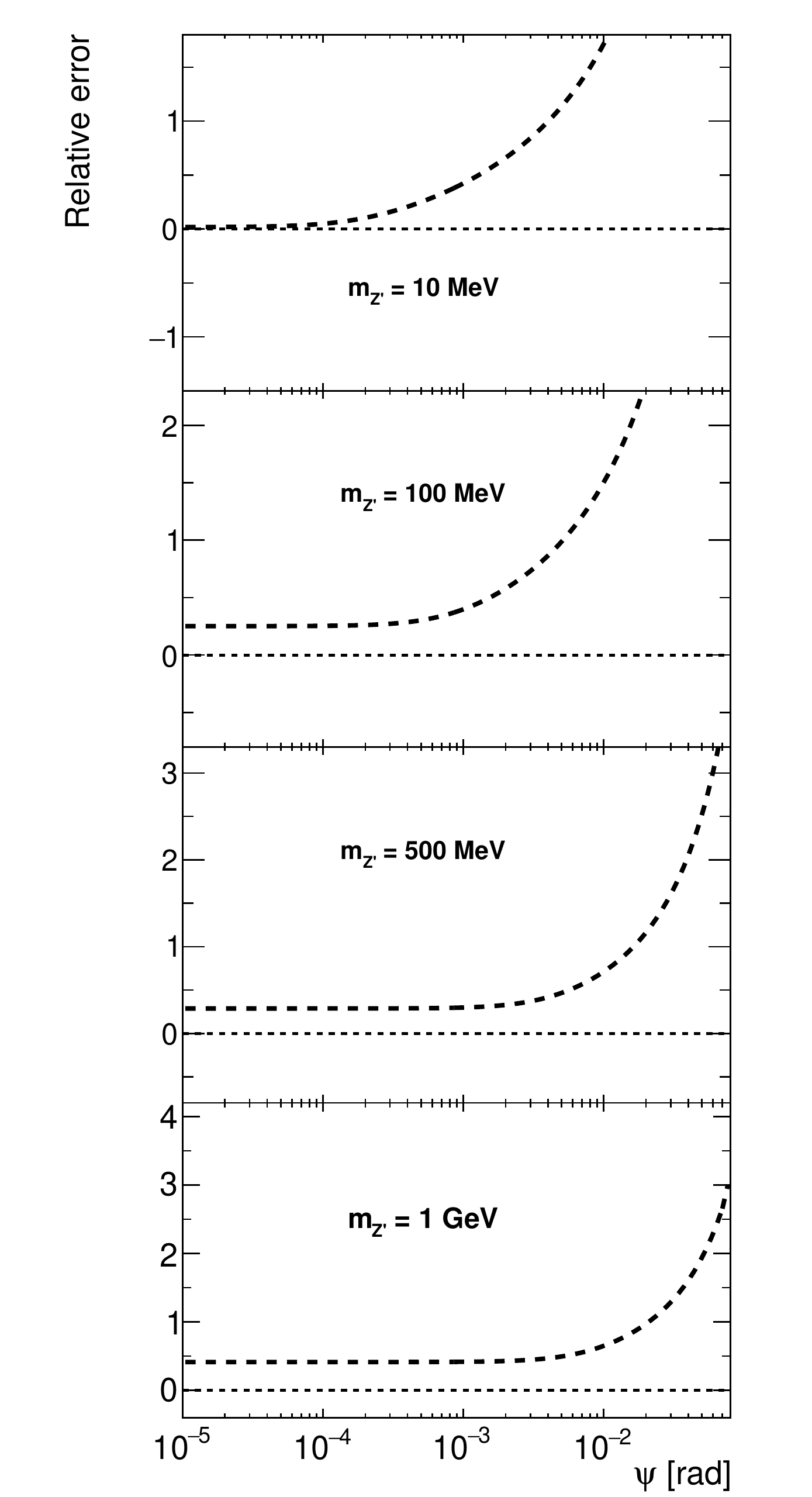}
    \caption{({\it Left}) Differential cross-section as a function of $\psi_{\mu'}$ for IWW (blue dashed line) and WW (green line) for different $Z'$ masses. ({\it Right}) Relative error between IWW and WW defined as $(\mathcal{O}_{IWW}-\mathcal{O}_{WW})/\mathcal{O}_{WW}$.}
    \label{fig:dsdpsi}
\end{figure*}

\begin{figure*}[]
    \centering
    \includegraphics[width=0.45\textwidth]{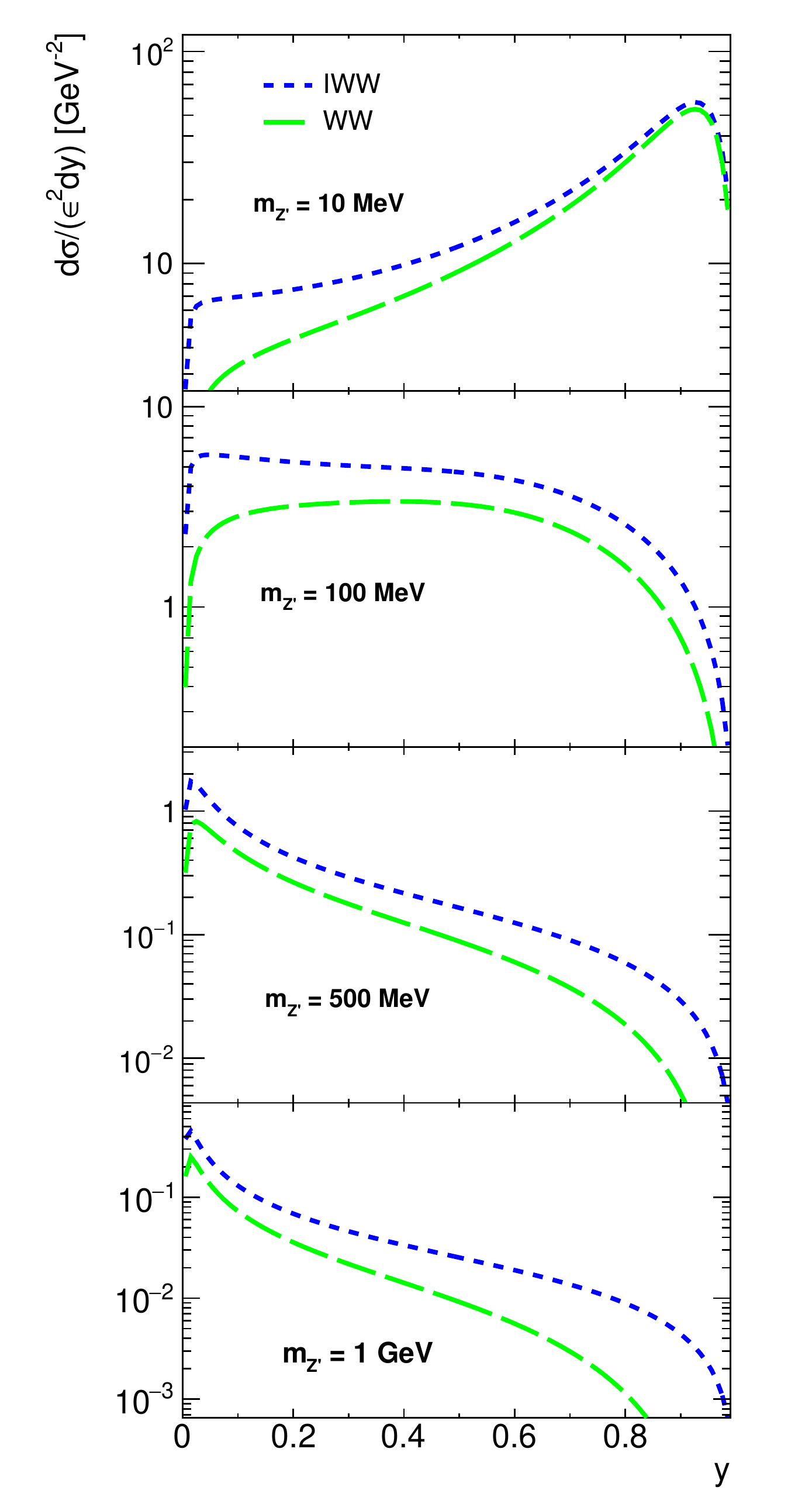}
    \hspace{5mm}
    \includegraphics[width=0.45\textwidth]{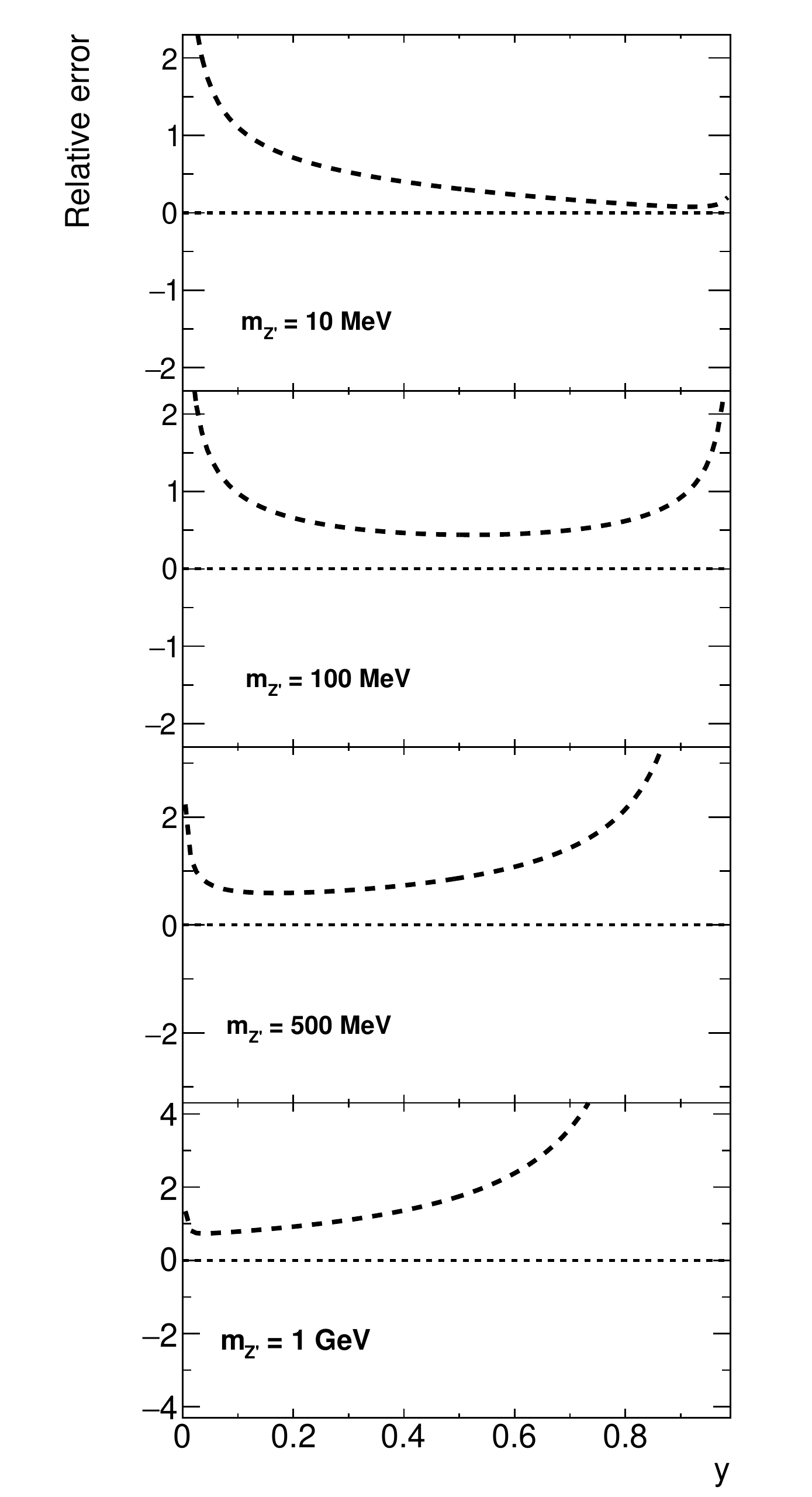}
    \caption{({\it Left}) Differential cross-section as a function of $y$ for IWW (blue dashed line) and WW (green line) for different $Z'$ masses. ({\it Right}) Relative error between IWW and WW defined as $(\mathcal{O}_{IWW}-\mathcal{O}_{WW})/\mathcal{O}_{WW}$. The angle of muon deflection is integrated in the range $\psi\in[0,0.1]$.}
    \label{fig:dsdy}
\end{figure*}

Fig. \ref{fig:dsdpsi} and \ref{fig:dsdy} show the results of the numerical 
integration of the double-differential cross-section with respect to the muon 
fractional energy and the recoil angle $\psi_{\mu'}$ for different $Z'$ masses in 
WW and IWW approximations. The difference between the two approximations becomes 
significant for large recoil muon angles and for large $y$ values due to the flux 
simplification used in IWW approximation described in the previous section. The 
behaviour of the differential cross-section as a function of $y$, reproduces what 
we observe as a function of $x$ taking into account that $y=1-x$ (see Fig. 
\ref{fig:dsdx_Z}).


\begin{figure*}[]
    \centering
    \includegraphics[width=0.45\textwidth]{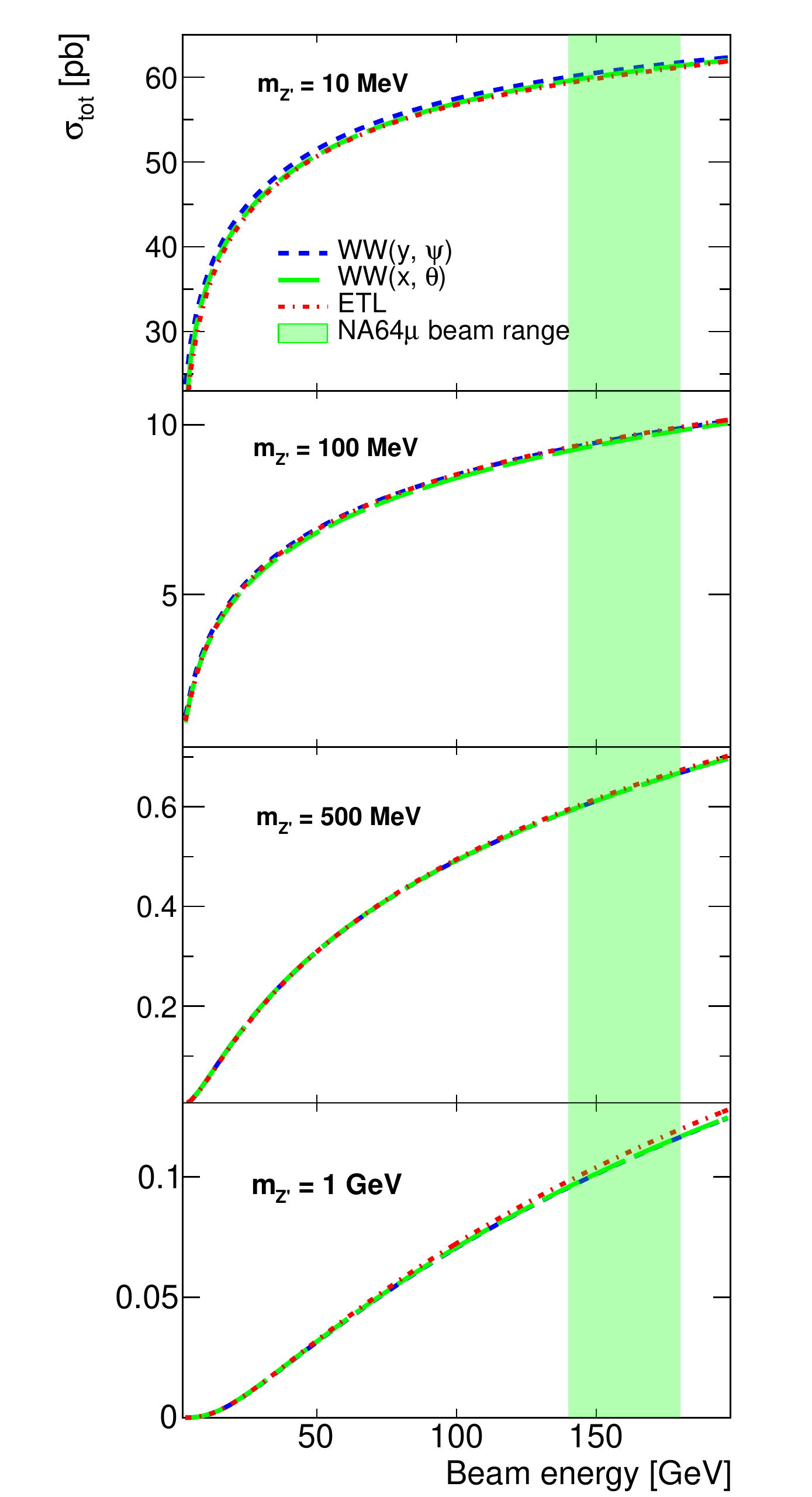}
    \hspace{5mm}
    \includegraphics[width=0.45\textwidth]{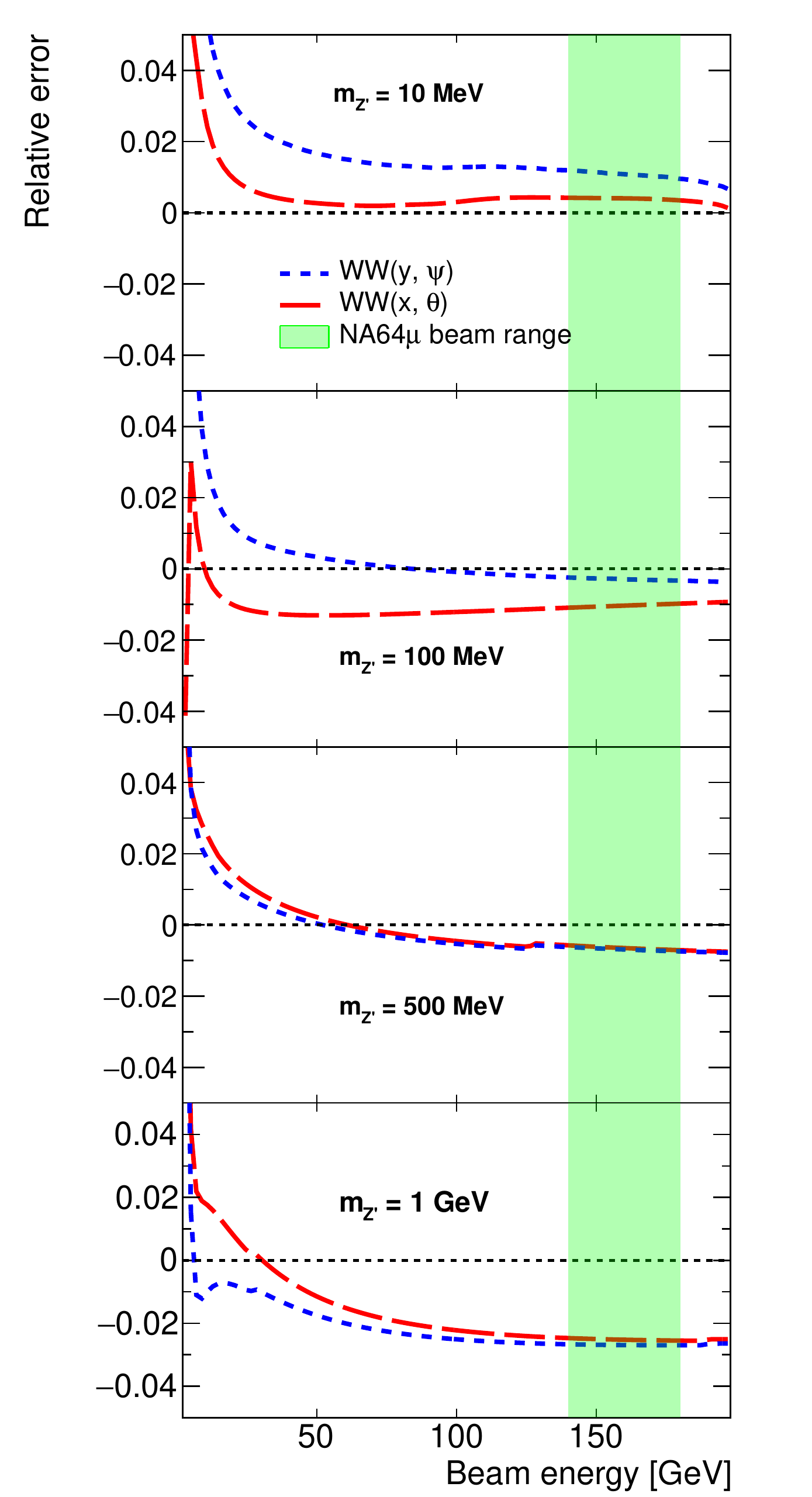}
    \caption{({\it Left}) Total cross-section as a function of the beam energy for WW expressions integrated over $y,\ \psi$ (blue 
    dashed line), $x,\ \theta$ (green line) and ETL integrated over $x,\ \theta$ (red dotted line).  
    ({\it Right}) Relative error between WW approximations and ETL as a function of the beam energy expressed as  
    $(\mathcal{O}_{WW}-\mathcal{O}_{exact})/\mathcal{O}_{exact}$. The typical NA64$\mu$ beam energy range is shown in green. The angles are integrated in the ranges $\theta\in [0,\pi]$ and 
    $\psi\in[0,\pi]$, the benchmark mixing strength is taken to be~$\epsilon=10^{-4}$.} 
    \label{fig:totWW_Z}
\end{figure*}

\begin{figure*}[]
    \centering
    \includegraphics[width=0.45\textwidth]{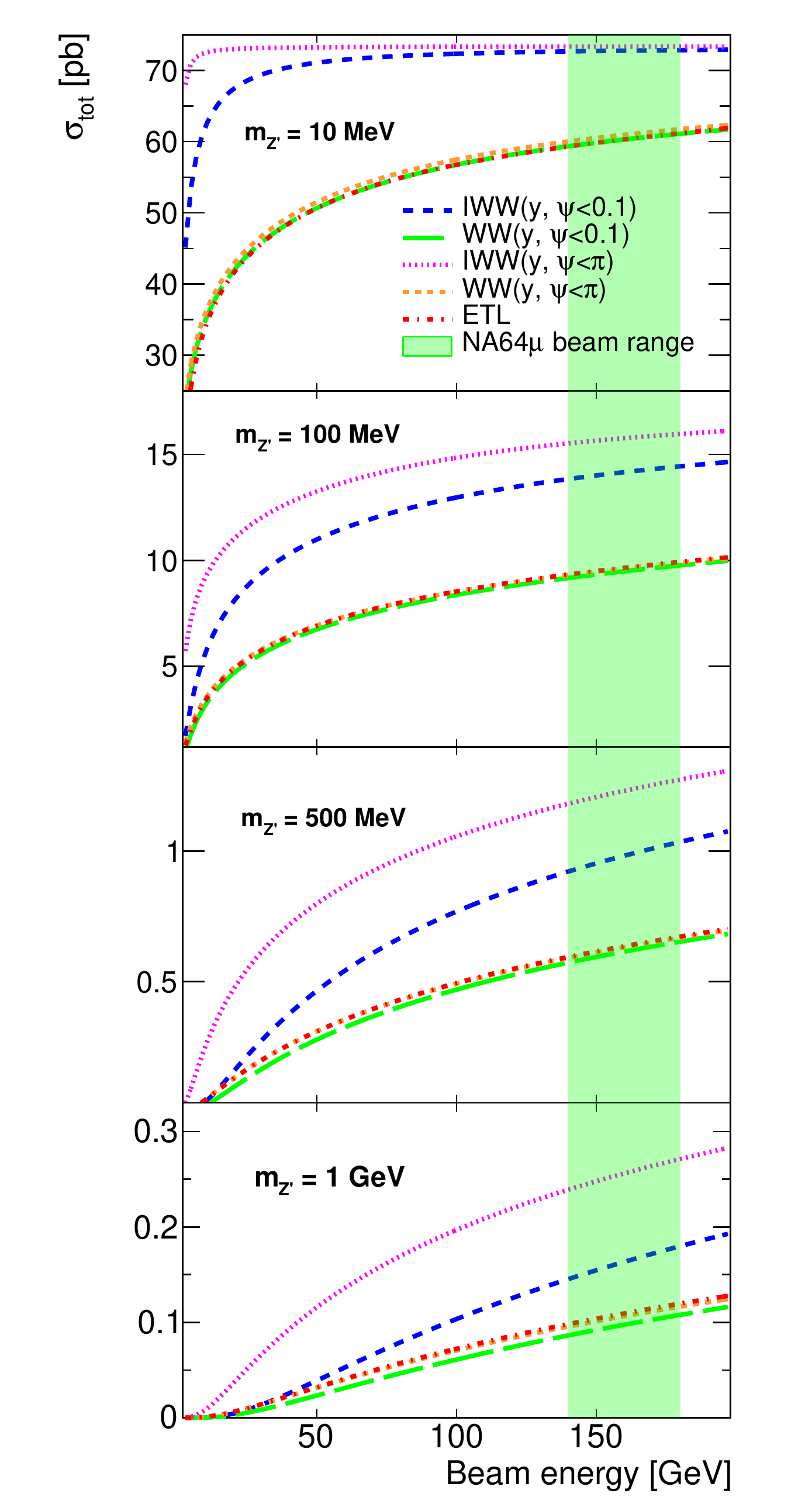}
    \hspace{5mm}
   \includegraphics[width=0.45\textwidth]{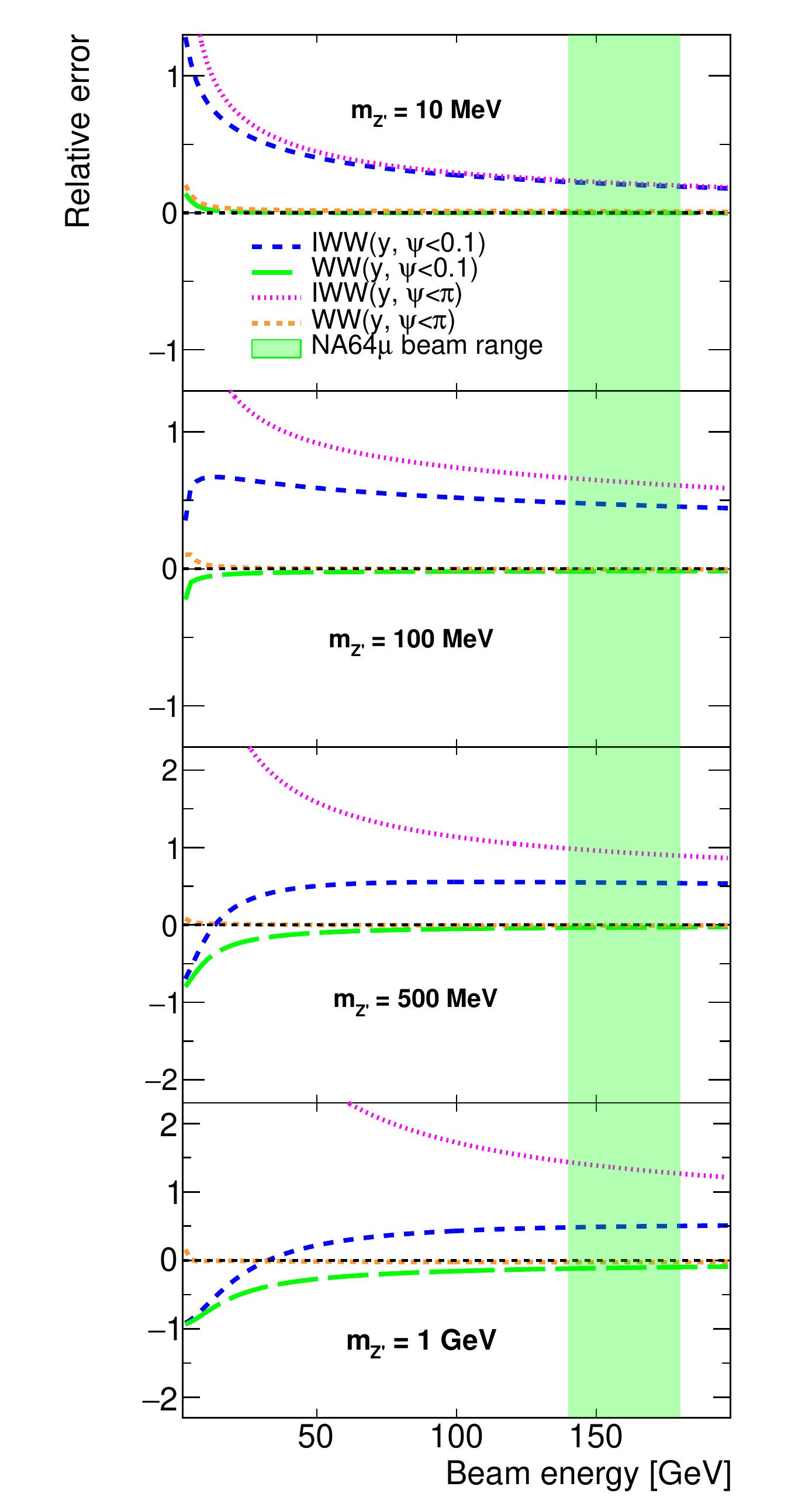}
    \caption{({\it Left}) Total cross-section as a function of the beam energy for IWW (blue/magenta dashed line), WW (green/orange line) expressions and ETL (red dotted line) integrated in the ranges  $0<\psi_\mu <0.1$ (blue and green) and $0<\psi_\mu <\pi$ (magenta and orange).  ({\it Right}) Relative error between WW, IWW approximations and ETL as a function of the beam energy expressed as  $(\mathcal{O}_{WW}-\mathcal{O}_{exact})/\mathcal{O}_{exact}$.  The typical NA64$\mu$ beam energy range is shown in green. The benchmark mixing strength is taken to be~$\epsilon=10^{-4}$.}
    \label{fig:totY_angle}
\end{figure*}

The total cross-sections in the  Weiszäcker-Williams approximation for both sets of 
variables, $(x, \theta_{Z'})$ and $(y,\psi_{\mu'})$ have been calculated as an independent 
cross-check. The results are compared in Fig. \ref{fig:totWW_Z} as a function of the 
beam energy. One can see from Fig.~\ref{fig:angular_spectra} that the typical
angle of $Z'$ emission is constant, it depends only on the energy of incoming muon $E_\mu\simeq 160$~GeV as $\theta_{Z'} \simeq
m_\mu/E_\mu$.
This implies that the total cross-section calculated for $Z'$ depends weakly on the angle cut $\theta^{max}_{Z'}$, as long as
$\theta^{max}_{Z'}\gg m_\mu/E_\mu \sim 4 \times 10^{-4}$. On the other hand, the
typical muon deflection angle is a function of the $Z'$ mass,
scaling as 
$\psi_{\mu'}\simeq m_{Z'}/E_{\mu}$ for a fixed incoming muon
energy. Thus, the total cross-section calculated for the deflected muon is sensitive to the maximum angle if 
$\psi^{max}_{\mu'} \gtrsim  m_{Z'}/E_{\mu}$. For this reason, in this case we 
integrate over the full parameter space of the muon and $Z'$ without restricting 
the outgoing particles angles. In particular, for Fig.~\ref{fig:totWW_Z} the 
integration ranges are 
$m_{Z'}/E_\mu<x<1-m_\mu/E_\mu$ and $0<\theta_{Z'}<\pi$ and $m_{\mu}/E_\mu<y<1-m_{Z'}/E_\mu$ and $0<\psi_{\mu'}<\pi$
for $Z'$ and the muon respectively.
The relative error of both cross-sections with respect to the exact cross-section 
integrated over $x$ and $\theta$ for all the masses considered here is below 2$\%$ for
energies above 10 GeV. \\ \indent  
The total cross-section derived integrating the muon fractional energy $y$ and the muon recoil angle $\psi_{\mu'}$ is
shown in Fig. \ref{fig:totY_angle} in IWW and WW approximations for both cases $\psi^{max}_{\mu'}=0.1$ and $\psi^{max}_{\mu'}=\pi$. 
The result is compared to the exact tree-level cross-section integrated over $x$ and $\theta_{Z'}$. 
Differences with respect to the exact values are negligible in the case of WW. However, in the low energy
region, especially for higher masses, the differences become significant. In order not to restrict regions the scattered muon cross-section has been integrated in a large recoil muon angle range ($\psi^{max}_{\mu'}=\pi$) (see magenta and orange dotted lines in Fig. \ref{fig:totY_angle}). The result is compatible with the one shown in Fig. \ref{fig:totX_Z}.

\section{Projected sensitivities to the mixing strength
\label{BoundsSection}}
\begin{figure*}
    \centering
    \includegraphics[width=0.48\textwidth]{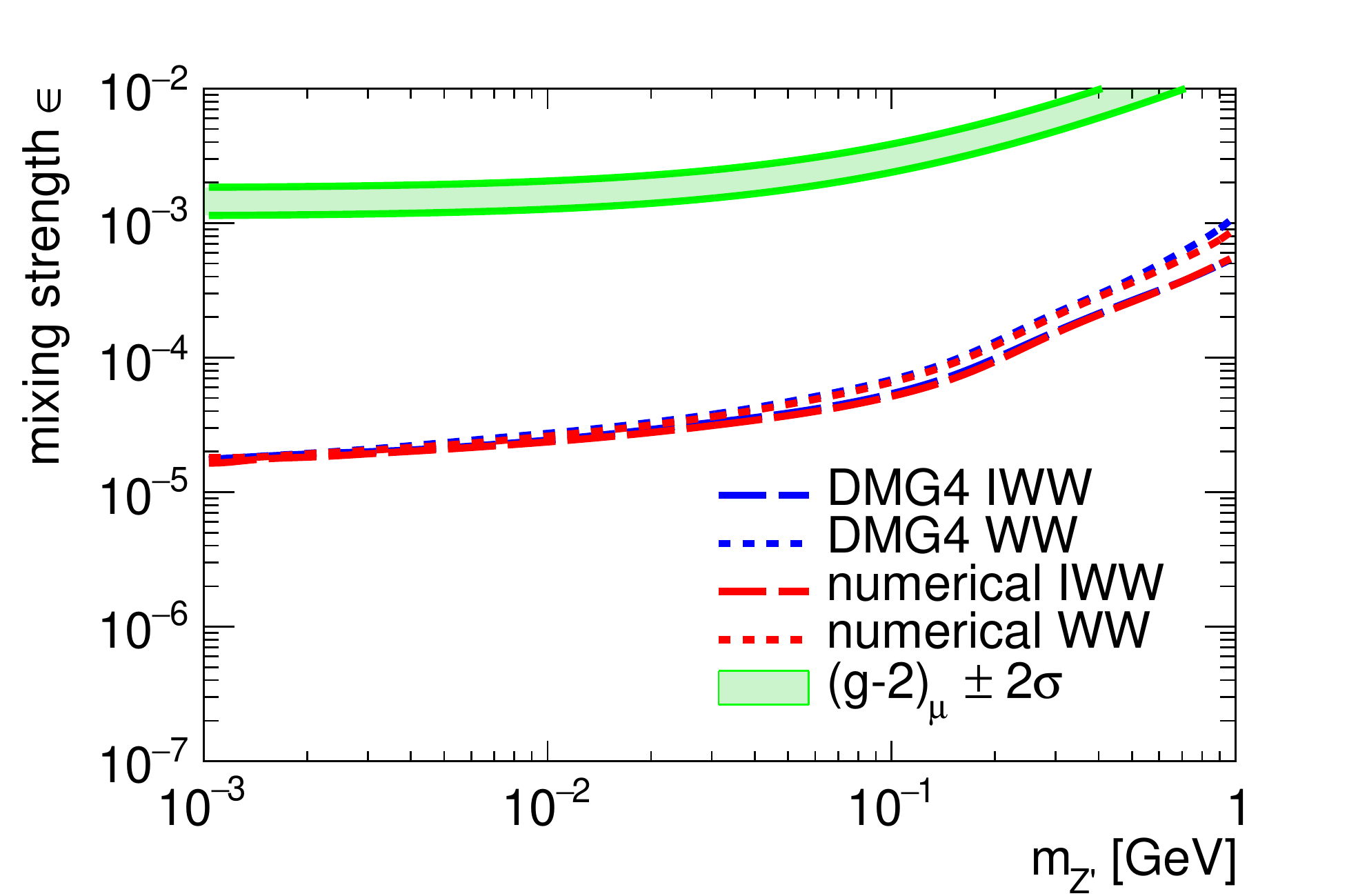}
    \hspace{0.5mm}
    \includegraphics[width=0.48\textwidth]{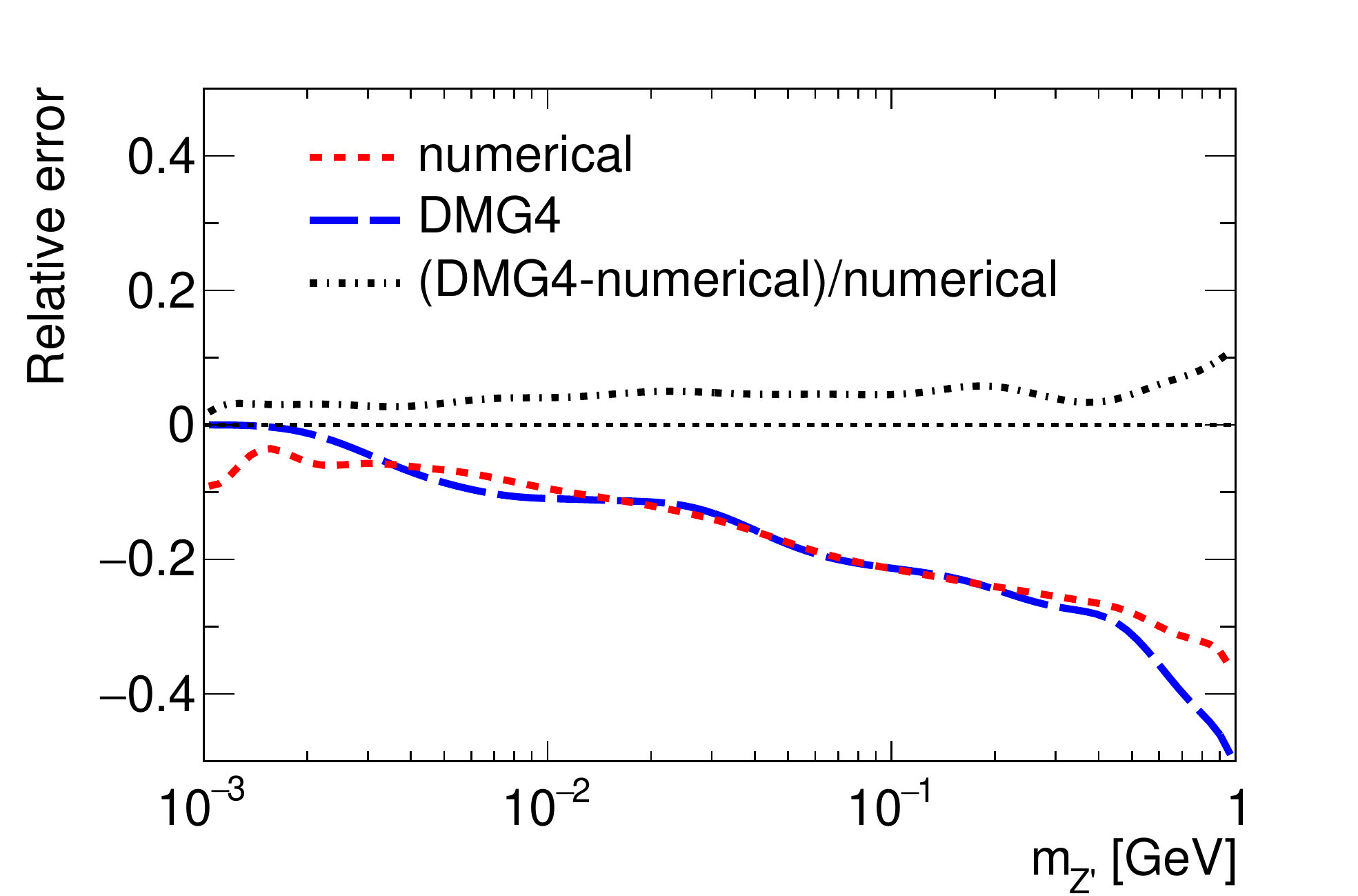}
    \caption{({\it Left}) Projected sensitivity in the $(m_{Z'},\ \epsilon=g'/\sqrt{4\pi\alpha})$ phase space obtained through both  MC simulations with DMG4 package \cite{Celentano:2021cna} (blue lines) and GSL \cite{Galassi} numerical integration (red lines) for $Z'$. The limits are calculated at 90\% C.L. for $10^{12}$ MOT and 160 GeV. Also shown is the $(g-2)_\mu$ favored band within $\pm2\sigma$. ({\it Right}) Relative error between IWW and WW defined as $(\mathcal{O}_{IWW}-\mathcal{O}_{WW})/\mathcal{O}_{WW}$. Is also shown the average relative error between MC simulations and numerical integration, defined as $(\mathcal{O}_{DMG4}-\mathcal{O}_{numerical})/\mathcal{O}_{numerical}$.}
    \label{fig:limits}
\end{figure*}

 The sensitivity of the experiment to $Z'$ is calculated according to \cite{na64mu}. In particular, the expected number of $Z'$, $N_{Z'}^{(\bar{\nu}\nu)}$ produced through muon bremsstrahlung and decaying invisibly to SM neutrinos, $Z'\rightarrow\bar{\nu}\nu$, is given by:
\begin{equation}
    \label{eq:sensitivity}
    \begin{split}
        N_{Z'}^{(\bar{\nu}\nu)}&=N_{MOT} \cdot \frac{\rho\mathcal{N}_A}{A}\cdot 
    \sum_i    \Delta L_{i}\cdot \sigma^{Z'}_{tot}(E_{\mu}^i) \cdot Br(Z'\rightarrow\bar{\nu}\nu) 
        \\
    \end{split}
\end{equation}
where  $A$ is the atomic weight, $\mathcal{N}_A$ is the Avogadro’s number, $N_{MOT}$
is the number of muons on target, $\rho$ is the target density,  $E_\mu^i$ is 
the  muon  energy  at  the  $i$th  step  in  the  target,  $\Delta L_i$ is  the  step 
length  of  the  muon  path  and  $\sigma^{Z'}_{tot}$
tot is  the  total  cross-section  of  the  $Z'$ production, $Br(Z'\rightarrow\bar{\nu}\nu)$ 
corresponds to  the branching ratio of $Z'$ decaying invisibly to SM neutrinos~\cite{Gninenko:2014pea}. \\ \indent
In the NA64$\mu$ facility it is assumed to utilize two, upstream and downstream, magnetic spectrometers allowing for precise measurements of momenta for incident
and recoiled muons, respectively~\cite{Gninenko:2014pea}. 
The muon missing energy signal of the reaction $\mu N\to \mu N Z', \, Z'\to \bar{\nu}\nu$ 
is defined by a scattered muon energy cut $E_{\mu'}\lesssim 80$~GeV.\\ \indent
The projected sensitivities are then calculated at 90\% C.L. 
for the number of signal events, i.e. it is required that $N_{Z'}^{(\bar{\nu}\nu)}>2.3$
events, assuming zero background. Fig. \ref{fig:limits} shows the projected sensitivities for both IWW and 
WW approaches in the plane $(m_{Z'},\ \epsilon)$ for $10^{12}$ 160 GeV muons on target (MOTs),
together with the values of $\epsilon$ necessary for the explanation of the $(g-2)_\mu$
anomaly. Those are obtained separately through numerical integration of Eq. 
(\ref{eq:sensitivity}) with GSL \cite{Galassi} and with robust MC simulations using the
DMG4 simulation package \cite{Celentano:2021cna}. It can be seen that in the low mass 
region, $m_{Z'}\lesssim10$ MeV, the relative error between IWW and WW does not exceed $10\%$, 
whereas it reaches about $40\%$ at masses $m_{Z'}\sim\mathcal{O}(1\ \text{GeV})$. It is
also worth noting that more conservative sensitivity lines are obtained with the 
realistic MC simulations, as a result of simulating the full muon physics within the
active target,
with about $\lesssim 10\%$ relative error with respect to 
the numerically integrated results.

\section{Summary}
In this work, we have derived, based on the work of \cite{Liu:2017htz}, the differential and total 
cross-sections for dark vector boson production in fixed target experiments through muon bremsstrahlung. We 
have shown that the commonly used improved Weiszäcker-Williams approximation differs significantly from 
the exact-tree-level calculations. On the other hand, the WW approach reproduces well the cross-section at a level of 
$<5\%$ in the high-energy beam regime, such as the one of NA64$\mu$ experiment. However, for very low beam 
energies, where $E_{\mu}\sim m_{Z'}$, the collinear regime is no longer valid and the phase-space 
approximations fail. We have also calculated the $Z’$ double-differential cross-section as a function of 
new variables, namely the scattered muon fractional energy and recoil angle, of particular importance for
Monte Carlo simulations and estimates in missing momentum experiments. Although the ETL for those variables
was not calculated, we have cross-checked our results against both WW approximation and ETL calculations as
a function of the emitted vector boson variables and found that the newly derived WW cross section reproduces the total 
cross-section with a good accuracy ($\lesssim 1\%)$. Additionally, we developed an analytical 
expression of the photon flux in WW approximation to reduce computational time due to numerical 
integration. Finally, our calculations were used to derive possible projected sensitivities in missing 
momentum experiments with both numerical integration and a full realistic Geant4-based MC simulation. It 
was found that in the high mass region, IWW calculations differ from WW ones, and thus ETL, by as much as 
40$\%$, over-estimating the sensitivity of muon beam fixed target experiments. Our results demonstrate the potential of these experiments to explore a broad coupling and mass parameter space region of dark vector bosons such as $Z'(A')$, including the very interesting muon $(g-2)$ unexplored region.

\section*{Acknowledgments}
We acknowledge the members of the NA64 collaboration for fruitful discussions, 
in particular, S. N. Gninenko, N. V. Krasnikov and E. Depero. The work of P. Crivelli, L. Molina Bueno and H. Sieber is supported by ETH Zürich and 
SNSF Grant No. 169133, 186181, 186158 and 197346 (Switzerland). The work of D. V. Kirpichnikov on MC simulation of $Z'$ emission is supported by the Russian Science Foundation RSF grant 21-12-00379.

 
\clearpage
\bibliographystyle{apsrev4-2}
\bibliography{../Bibliography/bibliographyNA64_inspiresFormat.bib,../Bibliography/bibliographyNA64exp_inspiresFormat.bib,../Bibliography/bibliographyOther_inspiresFormat.bib}

\end{document}